\documentclass[aps,pre,twocolumn,superscriptaddress,10pt]{revtex4-2}

\usepackage{graphicx,amssymb,amstext,amsmath,amsbsy,greekctr,color}
\usepackage{bbm}

\usepackage[utf8]{inputenc} % allow utf-8 input
\usepackage{url}            % simple URL typesetting
\usepackage{booktabs}       % professional-quality tables
\usepackage{amsfonts}       % blackboard math symbols
\usepackage{nicefrac}       % compact symbols for 1/2, etc.
\usepackage{lipsum}		% Can be removed after putting your text content
\usepackage{natbib}
\newcommand{\XP}{\mathbf{x}}       % path vector
\newcommand{\PP}{\mathcal{P}}      % TPS path probability
\newcommand{\DD}{\mathcal{D}}      % path differential
\newcommand{\ZZ}{\mathcal{Z}}

\newcommand{\PA}{\mathcal{P}_{\!A}}
\newcommand{\PAk}[1]{\mathcal{P}_{\!A,#1}}
\newcommand{\lmax}{\lambda_{\max}}
\newcommand{\D}{\mathcal{D}}
\newcommand{\intp}{\int\!\D\XP\;}
\begin{document}

\title{All reweighted paths lead to Rome: Unification of reweighting schemes for stratified path sampling}
%\title{All reweighted paths lead to Rome: comparison of path reweighting schemes for TIS}

\author{Peter G. Bolhuis*}
\affiliation{Van 't Hoff Institute for Molecular Sciences, Universiteit van Amsterdam,
%Science Park 904, 1098 XH 
Amsterdam, The Netherlands}
\email{p.gbolhuis@uva.nl}
\author{Rik S. Breebaart}
\affiliation{Van 't Hoff Institute for Molecular Sciences, Universiteit van Amsterdam,
%Science Park 904, 1098 XH 
Amsterdam, The Netherlands}

\author{Gianmarco Lazzeri}
\affiliation{Institute of Computer Science, Goethe University Frankfurt, Frankfurt, Germany}
\affiliation{Frankfurt Institute for Advanced Studies, Frankfurt, Germany}
\author{Jan-Hendrik Prinz}
\affiliation{keylight GmbH, Berlin, Germany}
\author{Roberto Covino}
\affiliation{Institute of Computer Science, Goethe University Frankfurt, Frankfurt, Germany}
\affiliation{Frankfurt Institute for Advanced Studies, Frankfurt, Germany}
\author{Titus S. van Erp }
\affiliation{Department of Chemistry and Biomedical Science,
Norwegian University of Science and Technology,
NO-7491 Trondheim, Norway
}

\date{\today}

\begin{abstract}
Path sampling-based stratification algorithms such as 
Transition interface sampling (TIS) enable the evaluation of kinetic rate constants and mechanisms of complex molecular rare event transitions. By sampling only short dynamical trajectories, TIS allows constructing  an estimate of the unbiased dynamics by reweighting the interface path ensembles. 
In the past decades several methods have been developed and applied to achieve such path reweighting, based on statistically optimal use of the trajectory data.
%dates. 
Why and how this reweighting exactly works has been underexplored up to now.  
 Here we  discuss several different schemes to compute    the reweighted path ensemble (RPE), notably  point-wise matching, Weighted Histogram Analysis Method (WHAM), Multistate Bennett Acceptance Ratio (MBAR),  variance based reweighting (VBR)  and  survival analysis reweighting (SAR).
 We show that except for the point-wise matching the other methods are leading to the same outcome, and in fact can be seen as identical. Moreover,  the SAR links to a well-established nonparametric maximum-likelihood estimator of product-limit form. This allows a synthesis of all path reweighting schemes, identifying the common underlying nature. All schemes follow from this common origin. 
 We illustrate the reweighting with some simple examples, and show that  care has to be taken when applying  WHAM, if the number of interfaces and histogram bins do not match.
 %does not match with the number of interfaces.    
\end{abstract}

\maketitle

\section{Introduction}

Molecular dynamics (MD) simulation has become a standard tool for investigating  complex molecular systems, and can predict both thermodynamic and kinetic properties, provided a sufficiently accurate force field is available.
Kinetic properties include the rate constants for rare but significant molecular transitions between metastable states, such as protein conformational changes, nucleation events, or chemical reactions. However, when dealing with high free energy barriers these rare transitions are interspersed with long waiting times in metastable states,  rendering the direct approach very inefficient.

 Numerous enhanced sampling techniques were developed to address this issue, usually by  biasing the system along a preselected collective variable (CV) that mimics the reaction coordinate\cite{Torrie_1974,Carter_1989,Huber_1994,Grubmuller_1995,Voter_1997,Laio_2002,Darve2001, Gao2008}.
The success of these  methods depends on the choice of an appropriate  CV, and  failure to do so can result in inaccurate estimates of barrier heights, rate constants, and even mechanistic details\cite{Chandler1998}.  Moreover, the biasing  approach distorts the natural dynamics itself.  
These challenges led to the development of path-based techniques that optimize 
%dynamic 
trajectories connecting reactant and product states, and enabling the identification of the minimum (free) energy pathway between them.
Prominent examples include Nudged Elastic Band\cite{Henkelman2000,Henkelman2000a,Henkelman2017}, string method\cite{E2005,E2010}, action minimization\cite{Lee2017}, and dominant reaction pathway\cite{Faccioli2010}. 

In the case of a  complex rough energy landscape at finite temperature, a more effective strategy is to sample from a distribution of dynamical paths\cite{Chong2017}.  
Transition path sampling\cite{Dellago1997,Dellago1999,Bolhuis2002,Dellago2002,Dellago2009,Bolhuis2009a,BolhuisDellago2015,Bolhuis2021} achieves this  by  collecting an ensemble of pathways that connect initial with final state, through performing a Markov Chain Monte Carlo random walk  through trajectory space.  
Starting with an initial trajectory, TPS generates a new trial path by selecting a frame and initiating a new trajectory by integrating the equations of motion both forward and backward in time\cite{Dellago1998,Bolhuis1998}. In a Metropolis-Hastings acceptance rule  based on  detailed balance, the trial path is then either accepted or 
rejected, and is  rejected if it fails to connect the two stable states.
%accepted/rejected. 
Repeating this process results in the transition path ensemble, providing unbiased insight into the mechanism of the rare molecular  transition.

By evaluating the reversible work between the reactive transition path ensemble and an ensemble of all trajectories  originating in the initial state, TPS also allows for computation of the rate constant. However, efficiency was greatly enhanced with the advent of transition interface sampling (TIS), which samples  a series of path ensembles conditioned to cross a  hypersurface, or interface,  positioned between the stable states\cite{vanErp2003,vanErp2005,vanErp2012}. 
Central to TIS is the crossing probability,
the probability that a trajectory crossing one interface will also cross the subsequent interface. The final interface then determines the probability of transitioning into the final state.  Multiplying the crossing probabilities of all interfaces with the flux through the initial interface enables precise calculation of rate constants. 
By stratifying the problem in this manner, TIS avoids the exponential increase in sampling cost with barrier height that is characteristic of direct simulations.
Each interface is defined by an order parameter or collective variable, typically represented with the variable $\lambda$. While TIS theoretically provides exact rate constants regardless of the chosen order parameter, 
this choice does affect the sampling efficiency, although the resulting loss in efficiency due to a suboptimal order parameter is generally less pronounced than that observed in many other sampling methods\cite{vanErp2006, vanErp2012}.
The introduction of replica exchange  TIS (RETIS) and variants thereof improves these statistics, i.e. converges quicker, by enhancing the decorrelation between sampled paths, particularly for those paths near the stable states\cite{vanErp2007,Bolhuis2008,Cabriolu2017,zhang2024highly}.  

In the original TIS formulation \cite{vanErp2003}, the crossing probability between two consecutive interfaces was determined solely from the path ensemble conditioned on crossing the first interface of the pair. However, if a sufficiently large number of paths is sampled, this crossing probability can also be estimated from other path ensembles conditioned on %crossing
interfaces closer to the reactant state. Incorporating this additional information is desirable as it leads to a more accurate estimate of the overall crossing probability between the stable states.

Similar to how biased configurational samples  in enhanced sampling techniques like umbrella sampling\cite{Torrie_1974} or metadynamics\cite{Laio_2002} can be unbiased by reweighting, entire TIS trajectories can also be reweighted to  correct for their overabundance, by ensuring that the reweighted trajectories matches the crossing probability statistics. 
This notion led to the concept of the reweighted path ensemble (RPE)\cite{Rogal2010}, which assigns to each sampled path in the TIS simulation a weight such that the resulting weighted ensemble provides a representative description of the trajectory statistics that would be obtained from exhaustive unbiased direct MD. 
This probability, or weight factor, turns out to be determined by the crossing probability at the maximum order parameter reached by the path and can be obtained
using reweighting schemes such as the Weighted Histogram Analysis Method (WHAM)\cite{Kumar1992}, Multistate Bennett Acceptance Ratio (MBAR)\cite{Bartels2000,shirtsStatisticallyOptimalAnalysis2008}, or, alternatively,   analysis based on expected variance contributions\cite{vanErp2016}, or the ``survival analysis reweighting" (SAR)\cite{Lazzeri2025archive}. 
Weighted averages computed on the RPE trajectories approximate equilibrium expectations, including, but  not limited to, free energy projections and rates.

Determining the crossing probability using % Transition Interface Sampling (TIS) 
TIS thus leads to  a proper reweighting of the sampled trajectories within the path ensemble as if they were from an exhaustive unbiased run. 
The key advantage is that the  RPE provides  more information %than just the rate constant. 
on paths near the top of the barrier or transition state compared to an unbiased path ensemble, given the same amount of total molecular dynamics (MD) effort.
This means that the RPE not only gives better estimates of the rate constants, but also on the free energy,  mechanistic information and other kinetic properties. 
Specifically, the RPE 
can give arbitrary free energy projections, and   estimates of the projected committor, as  demonstrated in Refs. \cite{Rogal2010,Lechner2010}.
In the subsequent years, the RPE has been  computed for various systems, e.g.  studies of crystallization\cite{Lechner2011,DiazLeines2018,Arjun2020}, chemical reactions\cite{vanErp2016},   protein folding\cite{Lazzeri2023}, kinetic optimization\cite{Bolhuis2023,Bolhuis2025}, transmembrane dimerization\cite{Jackel2025} and membrane permeation\cite{Safaei2025,vanErp2026}.

Interest in the  RPE has recently been revitalized with the introduction of the AIMMD algorithm\cite{Jung2023}. This algorithm efficiently samples the Transition Path Sampling (TPS) ensemble for complex molecular transition using machine-learned committor models, which also provide valuable mechanistic insights.  %Lazzeri et al.~complement 
AIMMD  can be combined with regular MD to reconstruct the entire RPE\cite{Lazzeri2023,Jackel2025}, or 
%with 
%. Recently, %Breebaart et al.~combined AIMMD  
with TIS to obtain accurate committor models based the entire RPE for high barriers~\cite{Breebaart2026understanding,Breebaart2026training}.

While the concept of the RPE is by now already 15 years old, it still remains a bit mysterious why and how this reweighting works in the first place. Moreover, the different methods for obtaining the weights for proper reweighting  differ slightly. 
In this paper we compare the different  approaches existing in the literature and reanalyze how they relate. We find that in fact all the different methods lead to identical  results, although the original WHAM approach exhibits  some inconsistencies when the histogram bins do not match with the  TIS interfaces.
These identities lead to a reformulation of the underlying foundation for the path reweighting, which is connected to  a well-established nonparametric maximum-likelihood estimator of product-limit form\cite{Kaplan1958,LyndenBell1971,Woodroofe1985}. 

The remainder of this paper is organized as follows. We start with an brief overview of TPS and TIS in section II. In section III we present the different methods for the reweighing appearing in the literature. 
Section VI provides the synthesis of the concepts and reformulates the foundation for the path reweighting. We then give some very simple numerical examples to illustrate how the path reweighting function in practice and indeed gives identical results. We end with a concluding discussion.

\section{Background}
\subsection{Transition path sampling}
 A trajectory  $\XP =  \{ x_0, x_1 \dots x_L\}$ is a sequence of  $L+1$ frames, where  each frame $x_i$  contains a full snapshot of the system at a time $i\Delta t$, with $\Delta t$ the time step between subsequent frames. The total duration of the path is thus $\mathcal{T} = L \Delta t$.
 Consider the distribution of free  paths
\begin{align}
    \PP [\XP] =  \rho(x_0) \prod_{i=0}^{L-1} p(x_i \rightarrow x_{i+1}) %/\mathcal{Z}. 
    \label{eq:freepathdist}
\end{align}
The first factor denotes the  (usual)  Boltzmann distribution,  $\rho(x) = \exp(-\beta H(x)) /Z$, with $H(x)$ the Hamiltonian or total energy, and $\beta =1 /k_BT$  the reciprocal temperature, with $k_B$ Boltzmann's constant, and $Z =\int dx \exp(-\beta H(x)) $. 
 The second factor is a product  over   $p(x_i \rightarrow x_{i+1})$,  the  short-time Markovian transition probability representing the underlying unbiased dynamics. This probability can be a delta function for deterministic dynamics, indicating that only one configuration $x_{i+1}$ follows from $x_i$ or a Gaussian function for stochastic dynamics\cite{Dellago2002,Bolhuis2009a}. Besides  the position of all particles, $x$ can also contain their momenta for deterministic or underdamped dynamics, and  additional information such as box-vectors.  
As all probabilities are properly normalized, 
the path integral is $\int \mathcal{D} \XP  \PP[\XP] =1$.

When we are interested in rare trajectories  like transition paths, a natural approach is to restrict the distribution that one samples. For a two-state system with an initial stable state A separated from  stable  state B  by a high barrier, one could be interested in the transition path ensemble of paths starting in A and ending in B. One can restrict the path distribution to only allow such paths by making use of  indicator functions $\mathbbm{1}_R (x)$,  which returns unity if the configuration is an element of the configuration space region R, and zero otherwise. The path distribution is then
 \begin{equation}
  \PP_{AB} [\XP] =  \mathbbm{1}_A (x_0) \mathbbm{1}_B (x_L)  \PP [\XP] /\mathcal{Z}_{AB},
 \end{equation}
 which is only non-zero for paths that start in state A at time $t=0$ and are in B at time $\mathcal{T} = L \Delta t$. 
 %Again 
 ${\mathcal Z}_{AB}$ is a normalizing factor  ensuring  
  $\int \mathcal{D} \XP  \PP_{AB}[\XP] =1$. 
Sampling from this distribution using a  Metropolis scheme 
is the focus of Transition Path Sampling (TPS)\cite{Dellago1997,Dellago2002,Bolhuis2009a,Dellago2009}.
A particular efficient implementation of TPS is the shooting move\cite{Dellago1998}. This algorithm first selects  a random, say the $k$th, frame $x_{k}$ from the current path $\XP$, named the shooting point.  The algorithm then generates a new path from $x_k$ by changing the momenta  (this is especially important for deterministic dynamics. For stochastic dynamics no modification is required, although sometimes useful\cite{Dellago2002,Bolhuis2009a,Dellago2009}), and then   by integrating the equation of motion by standard MD, Langevin, or Brownian dynamics or even MC forward in time until the $L$th frame. 
 Simultaneously a path is generated from the shooting point frame $k$,  backward in time until frame $t=0$ is reached. 
 This backward shooting is done by reversing the momenta and integrating forward in time, followed by a reversal of the momenta of all frames. The two partial paths, from $0$ to $k$, and from $k$ to $L$ constitutes the new trial path which can be accepted via the remarkable simple acceptance rule
 \begin{align}
     p_{acc}(\XP^{(o)}\rightarrow \XP^{(n)}) =\mathbbm{1}_A (x_0^{n}) \mathbbm{1}_B (x_L^{n}) \min \left[1, \frac{\rho (x^{n}_{sp})} { \rho(x^{o}_{sp})} \right].
     \label{eq:TPSaccrule}
 \end{align}
 where the  fraction of $\rho(x)$ in the min function accounts for any alteration of the shooting point\cite{Dellago2002,Bolhuis2009a,Dellago2009}.
This simple acceptance rule arises because the path is generated with the same dynamics as in the path distribution Eq.~\ref{eq:freepathdist}, causing a massive cancellation of factors\cite{Dellago2002,Bolhuis2009a,Dellago2009}.

\begin{figure}
    \centering
    \includegraphics[width=1\linewidth]{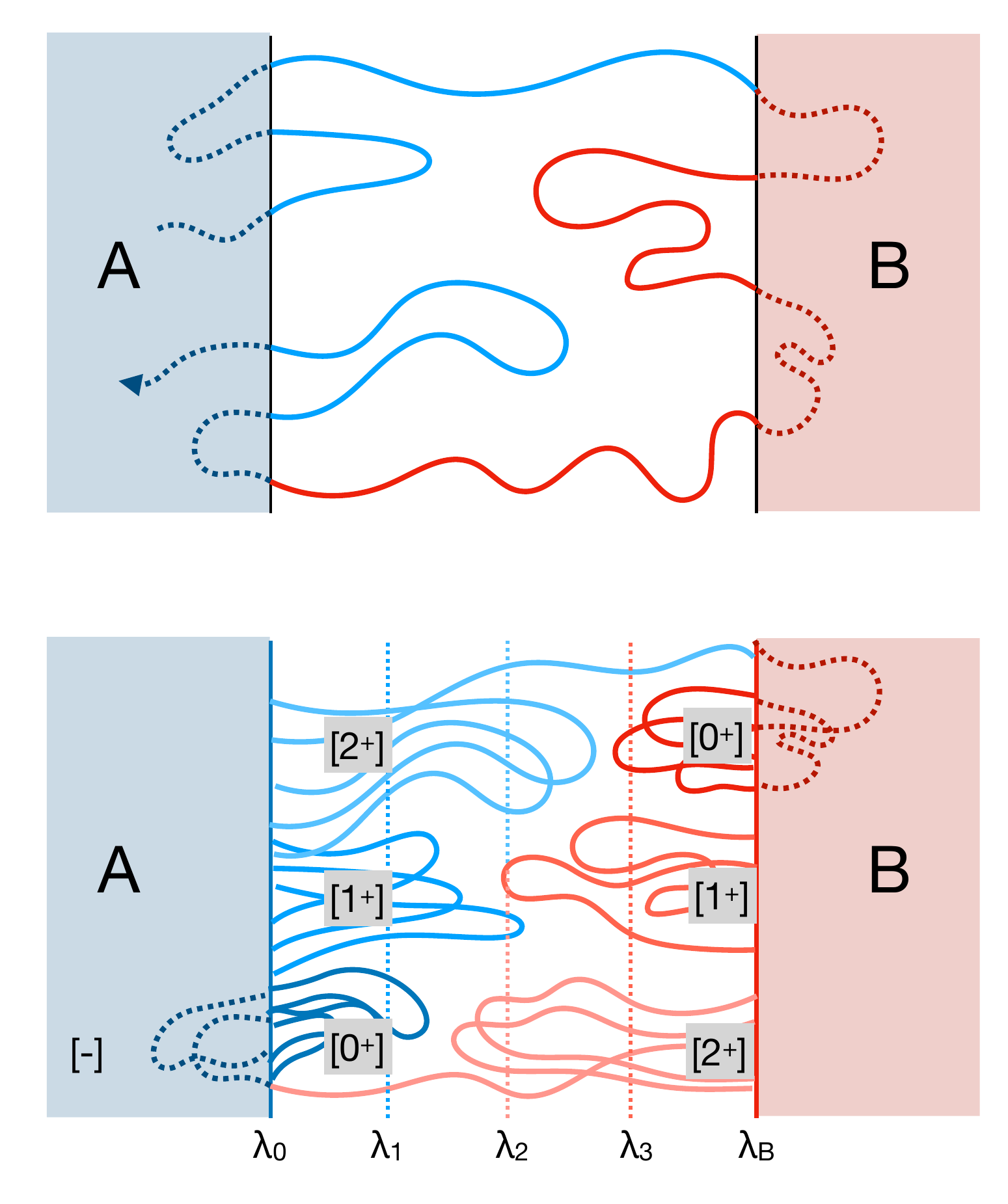}
    \caption{Top: an very long trajectory starting in A (colored blue) will make many attempts to cross the barrier, before hitting B, after which the subtrajectory is color red. Eventually the trajectory hits A again. The dotted sub-trajectories are the inA, inB paths. 
    %Straightforward MD is  inefficient.  
    Such a direct-MD trajectory is highly inefficient for sampling rare transitions, motivating the TIS approach shown below.
    Bottom: The way that TIS recasts   the problem into many interfaces ensembles with subtrajectories that cross their respective interfaces $\lambda_i$. }
    \label{fig:cartoonTIS}
\end{figure}
\subsection{Transition interface sampling}

Consider an infinite length dynamical trajectory exploring the space between stable state A and B. These states are  separated by a barrier region, which is unlikely to  be visited.  The trajectory frequently attempts to escape from A over the barrier, but many times returns to A before going to B. 
We can cut up this long path into many subtrajectories, resulting in several qualitatively different  path types. First there are the paths that are entirely in A, these paths start at the boundary of A explore A and end when it exits again. B has a similar type. These are usually called minus-interface paths in TIS, but we can also call them inA(inB) paths since they are always in A(B)~\cite{Lazzeri2023}.  
Then there are the AB or AA paths that exit A and either go on to B or return to enter A.
%These are the so-called AB or AA paths. 
Similarly for B we can define the BA and BB paths. See Fig.\ref{fig:cartoonTIS} for a schematic depiction.

The goal of TIS \cite{vanErp2003,vanErp2005,vanErp2012} is to sample all such paths efficiently and use them to compute the rate constant from the flux of trajectories out of the stable state A into B.
The rate constant is given by the following  expression
\begin{align}
    k_{AB} &= \phi_A P_A(\lambda_B|\lambda_A) \notag \\&= \phi_A \frac{\int 
    \DD \XP \PP[\XP] \mathbbm{1}_A (x_0) \mathbbm{1}_B (x_{L[\XP]})  }{ \int \DD \XP \PP[\XP] \mathbbm{1}_A (x_0)   },
    \label{eq:formalkAB}
\end{align}
where the first factor $\phi_A$ denotes the flux out of A through the boundary of A, and the second factor is the first crossing probability into B, given that the trajectory comes directly of A. In the second equality  this factor is identified as the fraction of all paths  leaving A that reach B for the first time before returning  to A. Here  we assume that the trajectories are halted as soon as they hit a stable state.  It is important to stress that this an exact equation, without approximation, other than that the process is a rare event, i.e. that when the trajectory hits B there is only a small 
chance to return to A\cite{vanErp2003,vanErp2005,vanErp2012}. 
Arguably, this is not even an approximation, since a well-defined rate constant already presupposes a separation of timescales.
A similar expression holds for the rate constant $k_{BA}$ connected to the reverse process. 

Direct evaluation of the second factor is very difficult, naturally, as  spontaneous transition paths will  occur very rarely. TIS solves this
%, just as the TPS rate calculation,
by a stratification approach.  While much more information can be found in  Refs.~\cite{vanErp2003,vanErp2005,vanErp2012}, here we give a  brief overview of the method. TIS defines a series of nonintersecting interfaces foliating the reactive space between the stables states. The interfaces are parameterized by a function  $\lambda(x)$, which monotonically increases with the interface index $i$. Standard TIS samples the path ensemble for each interface $\lambda_i$ where pathways start in the stable state A, cross the interface and then return to A or go on to stable state B. 
Thus, the TIS procedure creates AA and AB paths conditioned on crossing the interface $\lambda_i$. 
Similarly for the reverse process one can define a different set of interfaces for paths starting in B and crossing the interface from the opposite side. 

Performing path sampling on each of the
%one of the  set of
defined interfaces  leads to a set of $K$ conditional interface path ensembles and conditional crossing probabilities. From these conditional path ensemble one can construct the conditional crossing probability  $P_A(\lambda_{i+1}|\lambda_i)$, that is the conditional probability that a path coming directly from A and crossing interface $\lambda_i$ also crosses $\lambda_{i+1}$ before returning to A or going on to B.
The rate can then be approximated by multiplying all such conditional crossing probabilities
\begin{align}
    k_{AB} &= \phi_A P_A(\lambda_B|\lambda_A) =  \phi_A \prod_{i=0}^{K-1} P_A(\lambda_{i+1}|\lambda_i). 
\end{align}
While this is a correct way of performing TIS, one can make further progress by constructing a single crossing probability function 
$P_A(\lambda|\lambda_0)$ from the TIS path ensembles, which describes the unconditional probability for a path that leaves A and crosses some arbitrary interface defined at $\lambda$  
\begin{align}
    P_A(\lambda|\lambda_0) =\frac{\int 
    \DD \XP \PP_{A} [\XP] \theta(\lambda_{max}[\XP]-\lambda)}{ \int \DD \XP \PP_{A}[\XP]   },
\end{align}
where 
$\lambda_{max}$ returns the maximum value of $\lambda$ along the trajectory $\XP$, and the theta function equals unity if the argument is positive (non-negative) and zero otherwise. In other words, this restricted path ensemble only counts all  paths that cross $\lambda$. 
However, the TIS path ensembles only give the conditional probabilities 
\begin{align}
    P_A(\lambda|\lambda_k) =\frac{\int 
    \DD \XP \PP_{A,k} [\XP] 
    %\theta(\lambda-\lambda_k) 
    \theta(\lambda_{max}[\XP]-\lambda)}{ \int \DD \XP \PP_{A,k}[\XP]   },
\end{align}
where $\PP_{A,k}[\XP]$  denotes the path probability for the restricted interface ensemble $k$.
Given these conditional interface ensembles it is possible to reconstruct the full crossing probability by multiplying the path probabilities in the TIS ensemble with a reweighting factor. 

Since the crossing probability obey a very simple  rule\cite{vanErp2003}
\begin{align}
\label{eq:Pproduct}
       P_A(\lambda|\lambda_0) =   P_A(\lambda|\lambda_k) P_A(\lambda_k|\lambda_0),
\end{align}
it is clear that the multiplication reweighting  factor  for interface $k$ is the crossing probability $P_A(\lambda_k|\lambda_0)$ for that interface itself. 
Evaluating the  crossing probabilities for each interface sequentially would then give rise to  the weighting factor for the next interfaces, and eventually the full crossing probability function. This is the socalled point matching approach\cite{vanErp2003,vanErp2005}.
\begin{figure}
    \centering
    \includegraphics[width=1\linewidth]{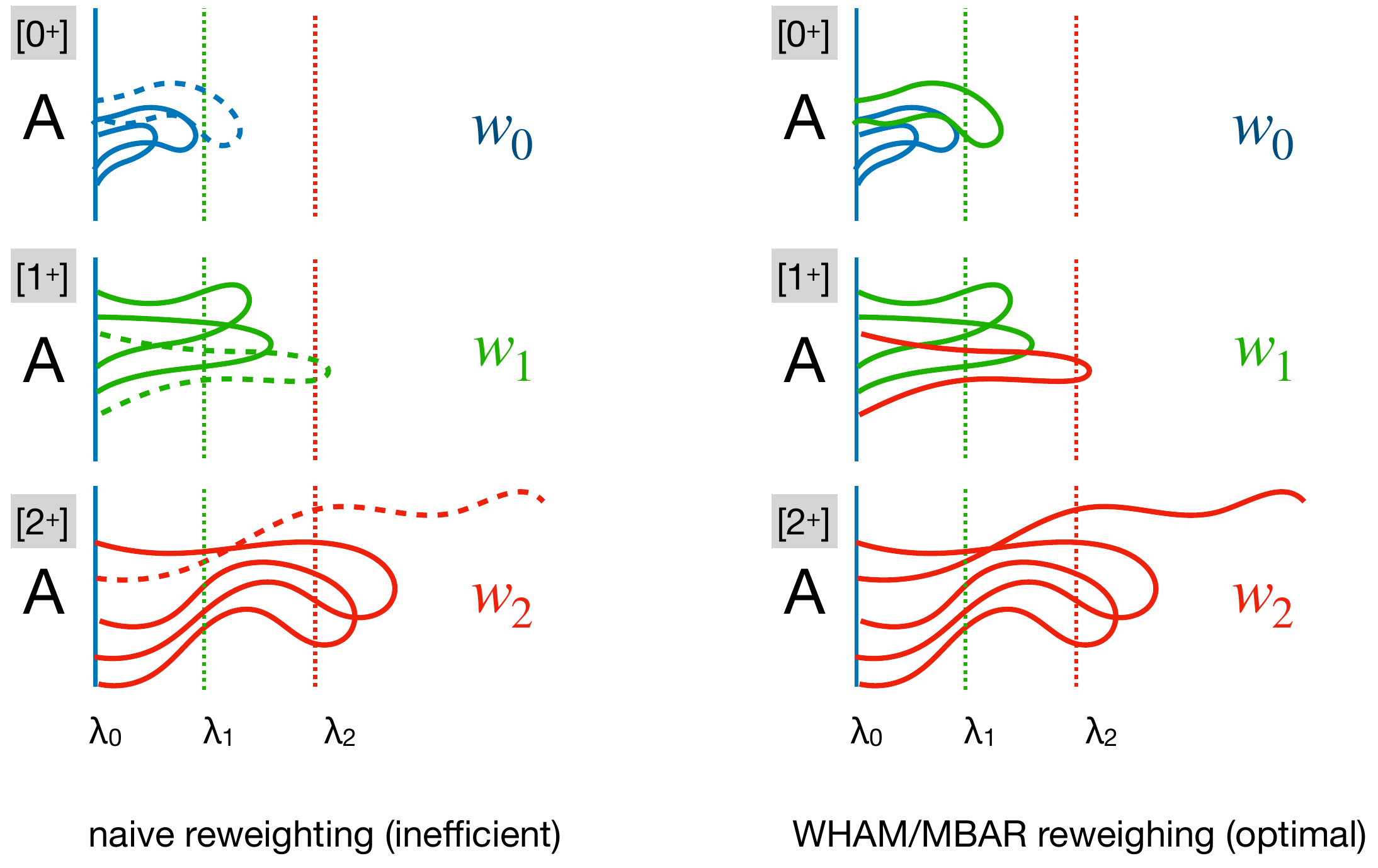}
    \caption{Left: assigning each TIS path ensemble a weight $w_i$, and discarding all trajectories that go beyond the next interface, is not  optimal  for reweighting. Right: The optimal reweighting assigns the same weight to each path that reaches $\lambda_i$, but not $\lambda_{i+1}$, for which no trajectories need to be discarded.  }
    \label{fig:correctreweighting}
\end{figure}
While the above has  been illustrated for the AB transition, the procedure for the BA transition is entirely analogous. 

For simplicity and clarity we focus here and in the remainder of the review/paper on the reactive region R between $\lambda_A=\lambda_0$ and $\lambda_B=\lambda_K$. 
Yet, the RPE can also be extended to include free simulations or minus-interface simulations\cite{Lazzeri2023}.

One consequence of the reweighting is that we do not only obtain the unconditional crossing probability, but also the unconditional path probability for the  interface ensembles, by  multiplying all paths probability by the crossing probability. This is the reweighted path ensemble, which provides a proxy of the unbiased ensemble $\PP_A[\XP]$,  and $\PP_{B}[\XP]$.

\subsection{Intuitive understanding of the RPE}

The RPE is a representation of the full path ensemble (restricted to region R) 
approximated by the sets of short dynamically unbiased paths obtained from  each TIS simulation, 
\begin{align}
    \{ (\XP_{k,1} , w_{k,1}), %(\XP_{k,2} , w_{k,2} ), 
    \dots, (\XP_{k,n} ,
    w_{k,n} ), \dots, (\XP_{k,N_k} ,w_{k,N_k} )\},
    \label{eq:PE}
\end{align}
where $k$ denotes the interface, and $w_{k,n} \equiv w[\XP_{k,n}]$ is the weight for the $n$th path in that interface ensemble. Consider now a collection of $K$ trajectory ensembles,
containing  $N_k$ samples.
The total path ensemble is thus approximated by
\begin{align}
    \PP_A[\XP] \approx \frac{1}{\mathcal{Z}} \sum_{k=0}^{K-1} \sum_n^{N_k}  w[\XP_{k,n}]  \delta[\XP-\XP_{k,n}].
    \label{eq:rpe}
\end{align}
with the  normalizing factor $\mathcal{Z}=\sum_k^{K-1} \sum_n^{N_k}  w[\XP_{k,n}]$.
The function $\delta[\mathbf{x} - \mathbf{x}_{k,n}]$ serves as an indicator function,
taking the value unity when $\mathbf{x} = \mathbf{x}_{k,n}$ and zero otherwise.
The RPE assigns weights to each sampled paths in the TIS simulations such that, for each (possibly time dependent) observable $O$ in the reactive space,
\begin{align}
\langle O \rangle =\frac{1}{\mathcal{Z}} {\sum_{k=0}^{K-1} \sum_n^{N_k}  w[\XP_{k,n}] O[\XP_{k,n}]}
    \label{eq:rpeav}
\end{align} 
in the limit of infinite sampling. That includes (but is not limited to) free energy projections.

As the weights should be directly related to the crossing probability, which is a monotonically decreasing function, the weights for each subsequent interface also should be smaller. Hence, the paths in later interfaces obtain a lower weighting factor than the ones closer to the stable state. This makes  sense since the former paths are oversampled compared to the latter.  So, while in an interface close to the transition state one path represents one path in the full path ensemble, in the low interfaces a single path can represent thousands or millions of paths. Since  straightforward sampling of the paths ensemble would only sample  these short excursions, this representation reduction allows the sampling  %makes it possible  to sample 
of transition paths in the first place.
In the following section we discuss different ways for computing these weights.

\section{Different roads towards the RPE}
\label{sec:RPE}

\subsection{Derivation of the RPE by point matching}

There are different ways to obtain the weights in Eq \ref{eq:PE}-\ref{eq:rpeav}.
Perhaps the simplest approach is to do point-matching, i.e.
we simply use 
\begin{align}
%w_k =  
P_A(\lambda_k|\lambda_0) =  \prod_{i=0}^{k-1} P_A(\lambda_{i+1}|\lambda_i)
\label{eq:pointmatching}
\end{align}
to assign the weight $w[\XP_{k,n}] = P_A(\lambda_k|\lambda_0)/N_k$ to every  of the $m_k$ paths $\XP_{k,n}$ that terminates between $\lambda_k$ and $\lambda_{k+1}$, while discarding the remaining $N_k-m_k$ paths that cross $\lambda_{k+1}$.
This is shown in the left panel of Fig.~\ref{fig:correctreweighting}.
However, this naive reweighting is statistically not optimal, and leads to a  higher uncertainty in the final corresponding probability as we have to discard many of the valuable sampled paths.
A  statistically more efficient approach is to retain and combine all available paths using 
% A  better way is to use 
an established optimal  reweighting scheme, such as WHAM or MBAR\cite{Kumar1992,Bartels2000,shirtsStatisticallyOptimalAnalysis2008}.

\subsection{Derivation of the RPE via WHAM}

The WHAM based reweighting was introduced in Ref\cite{Rogal2010}. Here we follow that argument but use a slightly more modern notation.
We start with the unnormalized crossing probability histograms obtained from the TIS simulations:
\begin{align}    
m_{A,k}(\lambda) = %\sum_k^{K-1} 
\theta( \lambda- \lambda_k)
\sum_n^{N_k} \theta( \lambda_{max} [\XP_{k,n}]- \lambda)  %\theta( \lambda- \lambda_k) 
\label{eq:crosproball}
\end{align}
where indices $k$ refer to the $K$ interface simulations, while $n$ runs over the $N_k$ samples in each TIS simulation.

The WHAM trick is to write the crossing probability as a reweighting of the 'sum of the unnormalized histograms'
\begin{align}
    P^{rw}_A(\lambda|\lambda_0) &= \sum_{k=0}^{K-1} w_k g_{k}(\lambda)
    %(\lambda)\theta(\lambda_{k+1}-\lambda) \theta (\lambda -\lambda_k )  
    \sum_{k'=0}^{K-1}
    m_{A,k'}(\lambda) \\&= 
    \sum_{k=0}^{K-1} w_k g_{k}(\lambda)%\theta(\lambda_{k+1}-\lambda) \theta (\lambda -\lambda_k )  
    m_A(\lambda),
    \label{eq:wham}
\end{align}
where  $g_{k}(\lambda) = \theta(\lambda_{k+1}-\lambda) \theta (\lambda -\lambda_k )$ %, so that the theta functions 
selects the correct weight for each 'interface bin' $\lambda_k \le \lambda < \lambda_{k+1}$.
In the second line we introduced the total sum of the unnormalized histograms  
\begin{align}
   m_A(\lambda)=  \sum_{k'=0}^{K-1}
    m_{A,k'}(\lambda),
\end{align}
where the sum can be extended to $K$ since  
%use to the fact that 
histograms do not contribute if $\lambda<\lambda_{k'}$.
The standard WHAM solution is to choose the interface weights such that the result is a consistent histogram, i.e. the reweighted histogram is the same as the unbiased histogram (see Eqs. A4-A7 in Ref.~\cite{Rogal2010}).
The optimal weights are then determined by minimizing the variance\cite{Kumar1992,FrenkelSmit}. This procedure leads to interface weights 
$w_i$
\begin{align}
w_i^{-1} = \sum_{k=0}^{i} \frac{N_k }{Z_k},
\label{eq:whamweights}
\end{align}
where the partition functions $Z_k$ turn out to be equal to the crossing probability values at the  interface $k$, $P_A(\lambda_k|\lambda_0)$, so that $Z_0=1$. The other $Z_k$ 
 follow from a set of self consistent equations\cite{Kumar1992}
\begin{align}
    Z_{k'} &= \sum_{i=0}^{M-1} \frac{ \sum_{k=0}^{K-1}m_{A,k}(\lambda^{(i)}) \theta(\lambda^{(i)}-\lambda_{k'}) }{\sum_{k=0}^{K-1} N_k \theta(\lambda^{(i)}-\lambda_k) /Z_k }\notag\\
   &= \sum_{i=0}^{M-1} \frac{ m_{A}(\lambda^{(i)}) \theta(\lambda^{(i)}-\lambda_{k'}) }{\sum_{k=0}^{K-1} N_k \theta(\lambda^{(i)}-\lambda_k) /Z_k } ,   
    \label{eq:whamselfconsist}
\end{align}
%(explained further in the SI), 
where $\lambda^{(i)}$ specifies one of the $M$ 
%is the  number of bins at $\lambda^{(i)}$ 
bins in the histogram. A convenient choice is $M=K$, so that every interface 'bin' $\lambda_k$ is also a bin at $\lambda^{(k)}$ in the unnormalized histogram.  These equations can be solved iteratively for ${Z}_k$.

Note that this procedure until now only reweighted the histograms, not the paths themselves.
The RPE can be obtained by realizing that paths that have their maximum $\lambda$ value in a particular interface bin, should be  treated identical in the WHAM, and should get the same weight. This is graphically illustrated in Fig.\ref{fig:correctreweighting}.
%This makes it clear that 
Therefore,
each path that has its maximum $\lambda$ value between $j$ and $j+1$ inherits a  weight 
\begin{align}
    w[\XP] =  \sum_{j} w_j g_j(\lambda_{max}[\XP]),
    \label{eq:pathweight}
\end{align} 
where the function $g_j(\lambda)$  now selects the correct WHAM weight based on how far the trajectory gets.

This means that the reweighted path ensemble can be written as a sum over all sampled paths, i.e., Eq.~\ref{eq:rpe},
and the crossing probability as
\begin{align}
    P_A(\lambda|\lambda_0)  &= \frac{1}{\mathcal{Z}} \sum_{k=0}^{K-1} \sum_n^{N_k}  w[\XP_{k,n}]   \theta( \lambda_{max} [\XP_{k,n}]- \lambda).
    \label{eq:RPEcrossprob}
   \end{align}
Thus,
all paths  for which  $\lambda_k< \lambda_{max} [\XP_{k,n}] % =  \Lambda 
\le  \lambda_{k+1}$, i.e., all paths that  reach $\lambda_k$, but not $\lambda_{k+1}$, obtain identical weights $w[\XP_{k,n}]=w(\lambda_{max}[\XP_{k,n}])\equiv w(\Lambda)$.
% Then it is clear that the only path $\bf x_i$ that survives the theta criteria, is the one that just reaches  $\Lambda$, i.e. $\lambda_{max}({\bf x}_i)  = \Lambda_i$]
Here, we introduced $\Lambda$ as a label to indicate how far a path reaches.
It follows that
\begin{align}
    P_A(\Lambda|\lambda_0)  = w[\XP_{k,n}] m_A(\Lambda) = w(\Lambda)m_A(\Lambda).
    \label{eq:GLseq}
\end{align}
%This equatin is central to the reweighting procesure 
%where we introdced 
This equation is used in Ref.~\cite{Lazzeri2025archive} to compute the weights for each path directly from the crossing probability. 

As the actual value  of the weight is dependent on  $\lambda_{max}$, we can  reorder the paths such that they are in order of $\Lambda$. Later we will see that this reordering is extremely convenient.

\begin{figure*}
    \centering
    \includegraphics[width=1\linewidth]{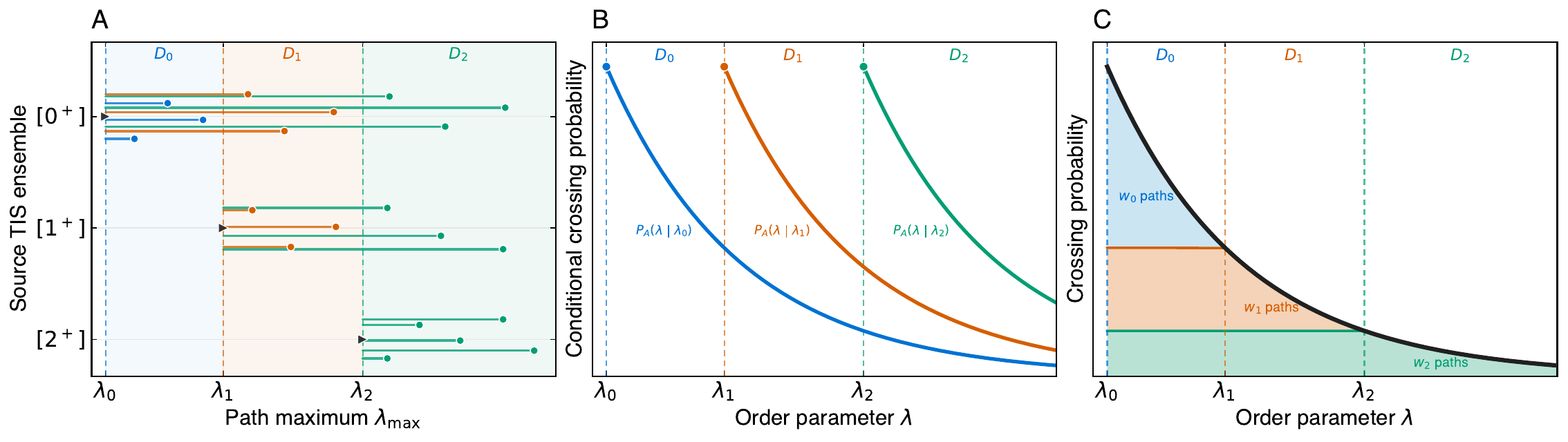}
    \caption{
    Illustration of path-wise WHAM/MBAR reweighting of TIS ensembles applied to crossing probability histograms. (A) Paths are colored according to the interface bin $D_i$ containing their maximum order parameter $\lambda_{\max}$, which determines their weight $w_i$. (B) The conditional crossing probabilities start at their respective interfaces and is computed for each individual path ensemble $[I^+]$. (C) The weighted contributions from all maximum-$\lambda$ bins combine to form a continuous cumulative crossing probability.
}
    \label{fig:WHAMexample}
\end{figure*}

\subsection{The MBAR Route}

Up to now we did not show how the optimal weights Eq.~\ref{eq:whamweights} and the related self-consistent equations Eq.~\ref{eq:whamselfconsist} themselves arise. We could derive these analogous to how WHAM is derived, by minimizing the expected error in the total crossing probability\cite{Kumar1992}, but in fact it is easier to derive the MBAR equations\cite{Breebaart2026combining}. 

Consider again a collection of $K$ trajectory ensembles,
% (biased) simulations, 
each of $N_k$ samples.
The unbiased probability $\PP[\XP]$ to observe a path $\XP$ based on this collection of sampled paths from A expressed as 
\begin{align}
    \PP_A[\XP] = \frac{1}{\mathcal{Z}} \sum_{k=0}^{K-1} \sum_n^{N_k}  w_{k,n} \delta[\XP-\XP_{k,n}].
    \label{eq:px}
\end{align}
Here, indices $k$ run over $K$ simulations, while $n$ runs over the $N_k$ samples in each interface simulation.  
The  factor $\mathcal{Z} = \sum_k \sum_n w_{k,n} $ normalizes the distribution. 
The weights of each path $w_{k,n}\equiv w[\XP_{k,n}]$ are unknown, but are also related to the biased path ensembles. For instance, in TIS we obtain biased path ensembles by imposing at each  interface a bias function ${h}_k[\XP] =\theta (\lambda_{max}[\XP] -\lambda_k )  $, which is only unity if the sampled path crosses interface $\lambda_k$.    
The distribution for such a biased ensemble is 
\begin{align}
    \PP_{A,k'}[\XP] = \frac{1}{\mathcal{Z}_{k'}}   \sum_{k=0}^{K-1} \sum_n^{N_k} w_{k,n} {h}_{k'}[\XP]  \delta[\XP-\XP_{k,n}],
    \label{eq:pk}
\end{align}
with  $\mathcal{Z}_{k'} =  \sum_{k=0}^{K-1}\sum_n^{N_k}  w_{k,n} {h}_{k'}[\XP_{k,n}]  $. Note that all paths from all simulations can contribute to this distribution. Thus for $k=0, \mathcal{Z}_0=\mathcal{Z}$, as all sampled paths cross the first interface.

In MBAR the unknown weights $w_{k,n}$ are determined by maximizing the likelihood to observe the sampled path sin their conditional ensemble. %by maximizing  the likelihood to observe the sampled paths. 
The log-likelihood is expressed as
\begin{align}
    \ln L = \ln \prod_{k=0}^{K-1} \prod_{n}^{N_k} \PP_{A,k}[\XP_{k,n}] = \sum_{k=0}^{K-1} \sum_{n}^{N_k}  \ln\PP_{A,k}[\XP_{k,n}]. 
\end{align}
Substituting $\PP_{A,k}[\XP_{k,n}] $ from Eq.\ref{eq:pk} gives
\begin{align}
    \ln L = \sum_{k=0}^{K-1} \sum_{n}^{N_k}  \ln  
    \frac{w_{k,n} {h}_k[\XP_{k,n}] }{\mathcal{Z}_k}.
\end{align}
Optimizing for  $w_{j,n}$ gives 
\begin{align}
    \frac{\partial \ln L}{\partial  w_{j,n} } = 0 = \frac{1}{w_{j,n}} - \sum_{k=0}^{K-1}  \frac{N_k  {h}_k[\XP_{j,n}]} {\mathcal{Z}_k}, 
\end{align}
where the factor $N_k$ derives from the summation of $n$.
%Dividing by $\mathcal{Z}$ and 
Rearranging gives 
\begin{align}
%    \frac{w_{j,n}}{\mathcal{Z}}  
w_{j,n}
    = \left[ \sum_{k=0}^{K-1}  \frac{N_k  {h}_k[\XP_{j,n}]} {\mathcal{Z}_k
    %\mathcal{Z}
    }\right]^{-1}.
    \label{eq:MBARweight}
\end{align}
Inserting this result in the definition of $\mathcal{Z}_k'$, Eq.~\ref{eq:pk}, yields 
\begin{align}
%    \frac{Z_{k'}}{\mathcal{Z}} = 
    %\sum_{j}^{K-1} 
    \mathcal{Z}_{k'}= 
   \sum_{j=0}^{K-1} \sum_{n}^{N_j}   \frac{{h}_{k'}[\XP_{j,n}]} { \sum_{k=0}^{K-1}  \frac{N_k  {h}_k[\XP_{j,n}]} {\mathcal{Z}_k
   %/ \mathcal{Z}
   }},
    \label{eq:MBARWHAMeq}
\end{align}
where  for clarity we replaced the outer $k$ index by $j$.
This is the implicit function that is central in MBAR, and which is identical to the central self-consistent equation in WHAM, Eq.~\ref{eq:whamselfconsist}. It can be solved iteratively for $\mathcal{Z}_k$, with one normalization constant fixed $(e.g., \ZZ=1)$, Eq.~\ref{eq:MBARweight} then yields the individual weights $w_{k,n}$ which in turn lead to the unbiased % We can also obtain the unbiased 
probability from Eq.~\ref{eq:px}
\begin{align}
    \PP_A[\XP] = \frac{1}{\ZZ} \sum_{j=0}^{K-1} \sum_n^{N_j} 
   \frac{  \delta[\XP-\XP_{j,n}]}  { \sum_{k=0}^{K-1}  \frac{N_k  {h}_k[\XP_{j,n}]} {\mathcal{Z}_k
   %/ \mathcal{Z}
   }
   }.
\end{align}
% WHEN n=1 -> w_k = ...
To make progress we first realize that for TIS ensembles the sum in the denominator only has contributions for paths where ${h}_k[\XP_{k,n}]$ is nonzero. Denoting the maximum interface that the path crosses $k_{max}$ (defined by $\lambda_{max}[\XP_{j,n}]$) this denominator reduces to 
\begin{align}
%\frac{w_{k,n}}{\ZZ} 
w_{j,n}=    \left( \sum_{k=0}^{k_{max}}  \frac{N_k %\mathcal{Z}
}  {\mathcal{Z}_k} \right)^{-1},
\end{align}
%This looks 
very similar to the WHAM expressions Eqs.~\ref{eq:whamweights}-\ref{eq:whamselfconsist}.
This means that the weight of each path is determined by the  maximum interface it crosses, as given by $\lambda_{max}[\XP_{j,n}]$. Indeed, this was the central result from the  treatment detailed in the previous section. 
The total crossing probability  can then be constructed from all complete reweighted paths in the RPE using
Eq.~\ref{eq:RPEcrossprob}.

Again, each path's weight is only dependent on the $\lambda_{max}[\XP_{j,n}]$. This makes it convenient to reorder the $N = \sum_k^{K-1} N_k$ paths in order of their maximum value $\Lambda \equiv  \lambda_{max}[\XP_{j,n}]$, and allows for the calculation of the weights to be simplified by looking at the number of paths that reach a certain bin or interface, the path number.

\subsection{Formulation of WHAM/MBAR in terms of path numbers}

The fact that each path  reaching interface $j$ but not $j+1$ receives a weight based on that bin, has lead to a simplified formulation in terms of path numbers \cite{Du2013,Newtonthesis}: $n_i(\lambda_j)$, the number of paths in interface ensemble $i$ reaching bin $j$. 
The conditional crossing probability for interface ensemble $i$ is then 
\begin{align}
    P_A(\lambda_j|\lambda_i ) = \frac{\sum_{k=j}^{K-1} n_i(\lambda_k)}{\sum_{k=i}^{K-1} n_i(\lambda_k)}.
\end{align}
Reweighting the path numbers then simply amounts to  a sum over all interfaces 
\begin{align}
    \tilde{n}(\lambda_k) = w_k  \sum_{j=0}^{K-1} n_j(\lambda_k), 
    %= w_k m_A(\lambda_k)
\end{align}
where the weight $w_k^{-1} =   \sum_{j=0}^{k-1} {N_j}/{Z_j}$ for each interface bin follows from solving the iterative WHAM/MBAR equations.
The reweighted crossing probability then becomes
\begin{align}
    P_A(\lambda_j|\lambda_0 ) = \frac{\sum_{k=j}^{K-1} \tilde{n}(\lambda_k)}{\sum_{k=0}^{K-1} \tilde{n}(\lambda_k)}.
    \label{eq:pathnumbers}
\end{align} 
This formulation can  be easily extended to multiple states, as was discussed in Ref\cite{Du2014,Newton2017b}.

\subsection{Evaluating  MBAR  in a single iteration}
The realization that  we
can reorder the sum  over the paths in Eq.~\ref{eq:MBARWHAMeq} without changing anything, leads to  a single sweep approach, that eliminates the need for an interactive solution of the WHAM/MBAR equations.
 If we order the paths according to $\lambda_{\max}[\XP_{j,n}]$, we can define the set of   paths $D_j$ that just cross the interface $j$, but not the next interface $j+1$ (see Fig.~\ref{fig:WHAMexample}).
Thus,
we have
\begin{align}
    %\frac{Z_{i}}{\mathcal{Z}} = 
    {Z}_i=
    %\sum_{j}^{K-1} 
   \sum_{j=0}^{K-1} \sum_{n\in D_j}   \frac{{h}_{i}[\XP_{j,n}]} { \sum_{k=0}^{K-1}  \frac{N_k  {h}_k[\XP_{j,n}]} { {Z}_k}}.
    \label{eq:MBARWHAMeq2}
\end{align}
Since the paths in $D_j$ do not cross $j+1$, the indicator function in the denominator 
%of path weight 
vanishes for interfaces above $j$, so $k <j$. That means that we can simplify the MBAR equation to  
\begin{align}
    %\frac{Z_{i}}{\mathcal{Z}} = 
    {Z}_i=
   \sum_{j=0}^{K-1} \sum_{n\in D_j}   \frac{{h}_{i}[\XP_{j,n}]} { \sum_{k=0}^j  \frac{N_k  } { {Z}_k}}.
    \label{eq:MBARWHAMeq3}
\end{align}
In the same way the indicator function in the numerator vanishes for $j<i$, whereas paths ending in interface $j$ are  clearly also crossing $i$, and therefore contribute. Since this is just counting the paths we end up with 
\begin{align}
   %\frac{Z_{i}}{\mathcal{Z}} = 
    {Z}_i=
   \sum_{j=i}^{K-1}   \frac{V_j} { \sum_{k=0}^j  \frac{N_k  } { {Z}_k}},
    \label{eq:MBARWHAMeq4}
\end{align}
where $V_j$ is the number of paths in $D_j$, i.e., those that cross $j$ but not  $j+1$, or in terms of path numbers $V_j= \sum_i n_i(\lambda_j)$.
We can write this equation %using Eq.~\ref{eq:whamweights}
\begin{align}
    %\frac{Z_{i}}{\mathcal{Z}} = 
    {Z}_i=
   \sum_{j=i}^{K-1} {w}_j V_j, 
    \label{eq:MBARWHAMeq5}
\end{align}
with   $w_j^{-1} =   \sum_{k=0}^{j} {N_k}/{Z_k}$ as before. 
This equation can be further reduced by considering the $Z$ for two subsequent interfaces $i$ and $i+1$. Since the expressions differ by just one term it is clear that
\begin{align}
    %\frac{Z_{i}}{\mathcal{Z}} = 
    {Z}_i -  {Z}_{i+1}  =  \sum_{j=i}^{K-1}  {w}_j V_j -
   \sum_{j=i+1}^{K-1}  {w}_j V_j  = {w}_i V_i. 
    \label{eq:MBARWHAMeq6}
\end{align}
This means that the MBAR weight for the next interface can be found without iteration, but simply by applying,  
\begin{align}
   %\frac{Z_{i}}{\mathcal{Z}} = 
    {Z}_{i+1} =   
    {Z}_i - {w}_i V_i =  {Z}_i - \frac{V_i }{\sum_{k=0}^i  \frac{N_k  } { {Z}_k}}, 
    \label{eq:Zip1}
\end{align}
starting from $i=0$. It is customary to set $Z_0=1$ for the initial interface. The values for the higher interfaces follow then straightforwardly, given the numbers $N_i$, and $V_i$, where the number $N_i$ denotes the paths belonging to interface $i$, 
while the number $V_j$ is the number of paths with their $\lambda_{\max}$ in between  the interface  $i$ and $i+1$.
In terms of path numbers,  $V_i= \sum_k n_k(\lambda_i)$, while $N_i= \sum_j n_i(\lambda_j)$.

\subsection{Variance based reweighting}

In Ref.\cite{vanErp2016} van Erp and coworkers arrive at the same equations starting from a different viewpoint.
They obtain the reweighted crossing probability from the path numbers, following a non-iterative single sweep 
\begin{align}
&P_A(\lambda_{k+1}|\lambda_0)    =
\frac{
\sum_{i=0}^{k} m_i(\lambda_{k+1})
}{
\sum_{j=0}^{k} N_j\left[
P_A(\lambda_j|\lambda_0)
\right]^{-1}
}.
\label{eq:titus}
\end{align}
where $m_i(\lambda_k) \equiv \sum_{j=k}^{K-1} n_i (\lambda_j) $ is the number of paths in interface i, that crosses $\lambda_k$ (note that $m_i(\lambda_i)=N_i$).
This expression  can be evaluated by iterating over $k=0,... K-1$ in a single sweep. 
Interestingly, the equation \ref{eq:titus} is derived using an argument based on the expected variance of the estimator\cite{vanErp2016, vanErp2026}. We briefly recap this argument below.

Suppose we seek the unbiased ensemble average $\left\langle O \right\rangle_0$ of observable $O$, but estimate it using a biased sampling distribution, yielding $\left\langle O \right\rangle_i$. If the effect of the bias can be accounted for by a scaling factor $b_i$, then
\begin{align}
\left\langle O \right\rangle_0
= b_i \left\langle O \right\rangle_i .
\label{eq:biasscaling}
\end{align}
Hence, given $K$ relevant biased ensembles in addition to the unbiased ensemble, we may combine the corresponding estimates as
\begin{align}
\left\langle O \right\rangle_0
= \frac{
\sum_{i=0}^{K-1} \alpha_i b_i \left\langle O \right\rangle_i
}{
\sum_{i=0}^{K-1} \alpha_i
},
\label{eq:alphabO}
\end{align}

While Eq.~\eqref{eq:alphabO} is mathematically exact for any choice of $\{\alpha_i\}$, it can be interpreted as a data-utilisation strategy when the ensemble averages are regarded as finite-sample estimates. In this interpretation, the above equation represents an approximate relation, where each $\left\langle O \right\rangle_i$ on the right-hand side denotes the numerical estimate obtained from a single simulation ensemble, with $i=0$ corresponding to the unbiased ensemble and $i>0$ to the biased ensembles. The left-hand side can then be viewed as a combined estimate of the unbiased ensemble average that makes use of the data from all available ensemble simulations.
The weights $\alpha_i$ are then chosen to minimise the statistical error of the combined estimate, thereby assigning greater weight to more precise estimates.

To make this more explicit, consider two simulations of different lengths that estimate the same quantity and are equally efficient. Since the statistical information is proportional to the simulation length, the weights should likewise be proportional to the respective simulation lengths. As the statistical error scales as the inverse square root of the simulation length, this is equivalently expressed as $\alpha_i \propto \epsilon_i^{-2}$, where $\epsilon_i$ denotes the relative statistical error of the single-ensemble estimate $b_i\left\langle O\right\rangle_i$. This weighting scheme allows simulations with different efficiencies to be combined and yields the same weights as obtained by minimizing the overall statistical error with respect to the $\alpha_i$, subject to the normalization condition $\sum_i\alpha_i=1$.

We now turn to the crossing probability, which incorporates both the path observable through $P_A(\lambda|\lambda_i)=\left\langle O(\mathbf{x})\right\rangle_i$, with $O(\mathbf{x})=\theta(\lambda_{\rm max}(\mathbf{x})-\lambda)$, and the scaling factors through $b_i=P_A(\lambda_i|\lambda_0)$. We then make approximations analogous to those commonly used in WHAM: we assume that the ensemble simulations have the same degree of correlation between samples (paths), neglect the statistical error in the scaling factors, and consider the regime where $\left\langle O(\mathbf{x})\right\rangle_i\ll1$.

To determine the statistical weights, consider first a Bernoulli observable that takes the value $1$ with probability $p$ and $0$ otherwise. Its ensemble average is $\langle O\rangle=p$, while the variance of a single sample is $\operatorname{Var}(O)=p(1-p)$. For $n$ independent samples, the variance of the sample average is 
$\operatorname{Var}\!\left(\overline{O}_n\right)
=p(1-p)/n$. The corresponding relative statistical error is therefore $\epsilon=\sqrt{(1-p)/(np)}$, giving $\epsilon^{-2}=np/(1-p)$. In the limit $p\ll1$, this reduces to $\epsilon^{-2}\approx np$.

Based on the relation $\alpha_i \propto \epsilon_i^{-2}$, we can apply the above result with $p=P_A(\lambda|\lambda_i)$ and $n=N_i/c$, where $c$ is the level of correlation between paths, i.e., the statistical inefficiency. We can neglect $c$, even when $c\gg1$, provided that it is the same for all ensemble simulations, since it only introduces an overall scaling of the $\alpha_i$ weights, which does not affect the final result.
In particular, for $c=1$, and choosing an overall scaling factor $1/P_A(\lambda|\lambda_0)$, we obtain (using Eq.~\ref{eq:Pproduct})
\begin{align}
\alpha_i
&=
\frac{N_i P_A(\lambda|\lambda_i)}
{P_A(\lambda|\lambda_0)}
=
\frac{N_i}{P_A(\lambda_i|\lambda_0)}
=
\frac{N_i}{b_i}.\label{eq:alphas}
\end{align}

Further we can write $N_i \left\langle O(\mathbf{x})\right\rangle_i=m_i(\lambda)$ 
with, as before, $m_i(\lambda)$ the number of paths in ensemble $i$ having $\lambda_{\max}(\XP)>\lambda$.
Applying that and Eq.\ref{eq:alphas} to Eq.~\ref{eq:alphabO} gives:
\begin{align}
&P_A(\lambda|\lambda_0)    =
\frac{
\sum_{i=0}^{k} m_i(\lambda
)
}{
\sum_{j=0}^{k} N_j\left[
P_A(\lambda_j|\lambda_0)
\right]^{-1}
}.
\label{eq:titus2}
\end{align}
with $k$ being the index that satisfies $\lambda_k < \lambda \leq \lambda_{k+1}.$ 
This means that if we consider $\lambda=\lambda_k$ we obtain Eq.~\ref{eq:titus}.

Notwithstanding the seemingly different form of the result we can identify the equality of the method to the  WHAM/MBAR  result by realizing the crossing probability is identical to $P_A(\lambda_k|\lambda_0) = Z_k/Z$, whith $Z=1$, so that  the weight in Eq. \ref{eq:whamweights} reads
\begin{align}
{w}_k^{-1} =
\sum_{j=0}^{k}
\frac{N_j }{Z_j}
=\sum_{j=0}^{k} N_j
[P_A(\lambda_j \mid \lambda_0)]^{-1}.
\label{eq:wP}
\end{align}
This means that Eq.\ref{eq:titus} becomes 
\begin{align}
    P_A(\lambda_{k+1}|\lambda_0)   = w_k   \sum_{i=0}^{k} m_i(\lambda_{k+1})  
    \label{eq:varbasedreweighting}
\end{align}
which, setting $\lambda=\lambda_{k+1}$, is indeed identical to Eq.~\ref{eq:wham}, where the unnormalized histograms are identical to $m_i(\lambda_{k})$.
Thus this variance  based reweighting (VBR) 
approach is identical to the WHAM/MBAR solution.  
Once the weights are established using  Eq.~\ref{eq:titus}, the RPE and a total crossing probability  can be obtained in the usual way by assigning a weight to each path using Eq. \ref{eq:pathweight}, and applying  Eq.~\ref{eq:rpe}  and Eq.~\ref{eq:RPEcrossprob}.

The above non-iterative solution is however clearly not (directly visibly) identical to the non-iterative solution Eq.~\ref{eq:Zip1} based on the path numbers, rather than the  unnormalized histograms. 
Yet, remarkably, they give exactly the same solution, as of course they  should, given that they both turn out to be identical to WHAM/MBAR.

\begin{figure*}[t]
    \centering
  \includegraphics[width=1\linewidth]{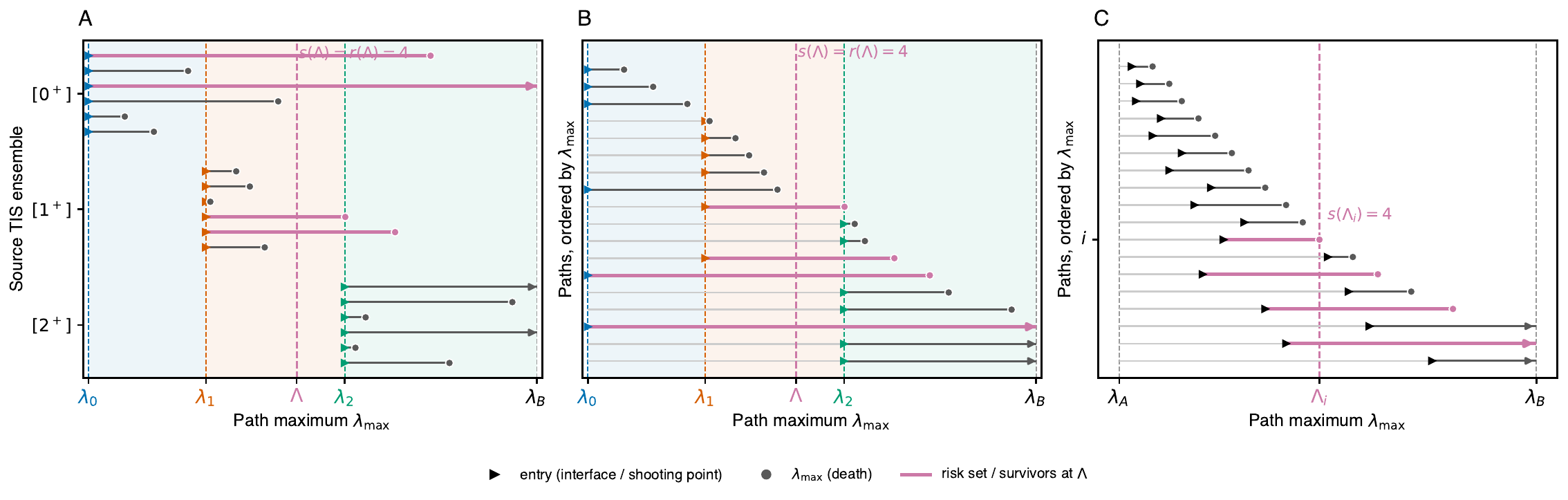}
    \caption{Schematic depiction of the SAR. 
    (A) Paths sampled in the TIS ensembles $[0^+]$, $[1^+]$, $[2^+]$. Each path enters at its ensemble’s interface (triangle) and dies at its maximum $\Lambda=\lambda_{max}$ value (dot). Reactive paths are halted (censored) at B (arrow).
    (B) The same paths, now ordered by $\Lambda$ regardless of their source ensemble. For a particular value $\Lambda$ (dashed pink line), the risk set contains the paths that have entered below $\Lambda$  and reach at least $\Lambda$  (thick pink); here ($s_A(\Lambda)=r(\Lambda)=4$. The number of still-active paths in the SAR, $s_A=S-F+V$, is therefore exactly the risk set $r$ of the product-limit estimator. Paths from higher ensembles are not yet counted, because below their interface they carry no information (left truncation). 
    (C) The limit of one path per interface, as in the rejection-free path sampling approach in Ref. \cite{Lazzeri2025archive}, where the shooting point of each path acts as its interface $\lambda_i$. Each path is then the only death at its own $\Lambda_i$, and the risk set $s_A(\Lambda_i)$ counts the paths that entered below $\Lambda_i$ and survive to it (here $4$).
In all panels the crossing probability $\hat S(\lambda)=P_A(\lambda|\lambda_0)$ drops by a factor $(r-1)/r$ at every path death. Because new paths enter only at interfaces, the product of these factors over all deaths within an interface bin reduces to the SAR drop Eq.~\ref{eq:drop2}. \label{fig:dropmethod}
}
%    (currently taken from Ref\cite{Lazzeri2025archive}} 
%    \label{fig:dropmethod}
\end{figure*}

\subsection{Survival Analysis Reweighting} %The "Drop Method"}

Another recently developed method for the crossing probability calculation, introduced in Ref\cite{Lazzeri2025archive}   amounts to computing the 'drop' in the crossing probability for each interface $i$.
This drop is the ratio  
\begin{align}
\frac{P_{\mathrm{A}}(\lambda_{i+1}|\lambda_0)} { P_{\mathrm{A}}(\lambda_{i}|\lambda_0)} = \frac{{Z}_{i+1}} {{Z}_i},
\end{align}
where we made use of the fact that for a given interface $i$ the crossing probability is proportional to  the partition function  $Z_i$. The ratio is also equal the fraction of paths crossing $i$ that  also cross   $i+1$, and thus 'survive'.
The survival analysis reweighting (SAR)   computes this ratio in a single sweep as one minus the fraction of paths that just  reaches its maximum $\lambda$ in between interface $i$ and $i+1$, with respect to  the number of 'still active' or 'surviving' paths $S_i - F_i + V_j$, thus 
\begin{align}
\frac{P_{\mathrm{A}}(\lambda_{i+1}|\lambda_0)} { P_{\mathrm{A}}(\lambda_{i}|\lambda_0)} =  1- \frac{V_i} {S_i - F_i + V_i}.
\label{eq:drop1}
\end{align}
Here $S_i=\sum_k^i N_k$ is the number of paths that have started in the interface ensembles up to $\lambda_i$
and $F_i=\sum_k^i V_k$ is the number that finished up to interface  $i$, i.e.  reach their maximum $\lambda$ at most at $\lambda_i$. The  additional $V_j$ in the denominator is needed as the number of paths that end between the $i$ and $i+1$ should be part of this fraction of surviving  paths.

While there are some heuristic arguments that this is a reasonable approach, we show here that SAR is in fact identical to solving the MBAR/WHAM equations. 
We start by dividing  Eq.~\eqref{eq:Zip1} by $ Z_i$ to obtain 
\begin{align}
   \frac{{Z}_{i+1}} {{Z}_i} = 1 -  
     \frac{{w}_i V_i }{{Z}_i}.
    \label{eq:Zif}
\end{align}
Thus, for the SAR to be  identical to the MBAR/WHAM result, we have to show that 
\begin{align}
 1- \frac{{w}_i V_i }{{Z}_i} \overset{?}{=} 1- \frac{V_i} {S_i - F_i + V_i}. 
\end{align}
Clearly, this means that we have  to show
\begin{align}
 \frac{{Z}_i}{{w}_i  } \overset{?}{=} {S_i - F_i + V_i} 
 \label{eq:simpledrop}
\end{align}
For $i=0$, we have $S_i = N_0$, $F_0=V_0$, and  ${w}_i = Z_0/N_0$, with $Z_0=1$. This gives for $i=0$
\begin{align}
 \frac{{Z}_0}{ Z_0 } N_0 =  N_0 - V_0 + V_0= N_0,
\end{align}
which is clearly always true.
We proceed by looking at the next interface $i+1$,  for which it should hold that 
\begin{align}
 \frac{{Z}_{i+1}}{{w}_{i+1}  } \overset{?}{=} S_{i+1} - F_{i+1} + V_{i+1}. 
\end{align}
Using the fact that $S_{i+1}= S_i + N_{i+1}$, $F_{i+1}= F_i + V_{i+1}$ and that, from Eq.~\ref{eq:whamweights}, 
\begin{align}
\frac{1}{w_{i+1}} = \frac{1}{w_i} + \frac{N_{i+1}}{{Z}_{i+1}}
\end{align}
%And that $S_{i+1}= S_i + N_{I+1}$
it follows, after substitution, that 
\begin{align}
 {{Z}_{i+1}} \left( \frac{1}{{w}_{i}} +  \frac{N_{i+1}}{{Z}_{i+1}} \right) =  S_{i} + N_{i+1} - F_{i} -V_{i+1} + V_{i+1}, \notag
\end{align}
which reduces to
\begin{align}
\frac{{{Z}_{i+1}}}{{w}_{i}}   =  S_{i}  - F_{i}. 
\end{align}
% Introducing the trivial factor $Z_i/Z_i$
% \begin{align}
% \frac{{{Z}_{i+1}}}{ Z_i}  \frac{ Z_i}{{w}_{i}}  =  S_{i}  - F_{i}, 
% \end{align}
% and realizing that   the first factor is given by Eq.~\eqref{eq:Zif}  we obtain
Substitution of Eq.~\eqref{eq:Zif} yields
\begin{align}
\left (1 -  
   \frac{{w}_i V_i }{{Z}_i} \right)  \frac{ Z_i}{{w}_{i}}   =  S_{i}  - F_{i}, 
\end{align}
which works out to be 
\begin{align}
 \frac{ Z_i}{{w}_{i}}  -V_i  =  S_{i}  - F_{i}, 
\end{align}
which is indeed  identical to  Eq.~\eqref{eq:simpledrop} that we set out to prove. If Eq.~\eqref{eq:simpledrop} holds for $i$, it also holds for $i+1$. Since we have shown it holds for $i=0$, it thus holds for all $0\le i \le K$ by induction. 

We can now obtain the working equation for the SAR from Eq.~\eqref{eq:drop1} by 
defining the number of surviving paths  $s_{\mathrm{A}}(\lambda_i ) = S_i - F_i  +V_i$, 
%as  the number of still active paths 
leading to 
\begin{align}
\frac{P_{\mathrm{A}}(\lambda_{i+1}|\lambda_0)} { P_{\mathrm{A}}(\lambda_{i}|\lambda_0)} =  \frac{s_{\mathrm{A}}(\lambda_i ) -V_i} {s_{\mathrm{A}}(\lambda_i )}. 
\label{eq:drop2}
\end{align}

The conclusion is that computing the crossing probability and weights by the SAR  method, is completely equivalent to solving WHAM/MBAR equation in a non-iterative way. Thus it can be argued that  SAR also uses  all information in the most efficient manner.

\subsection{The SAR for one path per interface}

The SAR was especially designed to be capable of handling the rejection free path sampling approach in Ref.~\cite{Lazzeri2025archive}.
Here, each interface is defined by the shooting point of a path, leading to, effectively, a single path per interface. (As a side note: In this treatment  each shot  (backward and forward) must be considered individually, so one full trajectory can potentially contribute with two drops).
The SAR takes the interface values $\lambda_{i+1}$ such that $V_i =1$, which is guaranteed by taking  the next interface only an vanishingly small $\epsilon$ further, ie. $\lambda_{i+1} = \lambda_i + \epsilon$, yielding 
%The results is thus 
\begin{align}
\frac{P_{\mathrm{A}}(\lambda_{i}+\epsilon|\lambda_0)} { P_{\mathrm{A}}(\lambda_{i}|\lambda_0)} =   \frac{ s_{\mathrm{A}}(\Lambda_i )-  1} {s_{\mathrm{A}}(\Lambda_i )},
\label{eq:drop3}
\end{align}
%which is indeed equivalent to Eq.~\eqref{eq:dropA}
for $\Lambda_i = \lambda_i$ (see Fig.~\ref{fig:dropmethod} for a graphical illustration). 
It is remarkable that this even works. 
However, this result can in fact be linked to  an exact
nonparametric maximum-likelihood estimator of product-limit form, a well-established result in the statistics literature\cite{Kaplan1958,LyndenBell1971,Woodroofe1985}.
%\subsection{Linking the SAR to Kaplan-Meijer}
The particular limit form of the SAR can be argued to  be a maximum likelihood estimate for independent excursions.
The individual contribution to the likelihood of interface $\lambda_i=\Lambda_i$ is 
\begin{equation}
    \mathcal L_{\mathrm{A},i} = p_i^{s_A(\Lambda_i)-1}  (1-p_i),
\end{equation}
where $p_i$ denotes the survival probability for a trajectory segment to go beyond the interface $\Lambda_i$. The total likelihood is the product over all  interfaces $\Lambda_i$
\begin{equation}
    \mathcal{L}_{\mathrm{A}}= \prod_i \mathcal L_{\mathrm{A},i} =\prod_i p_i^{s_A(\Lambda_i)-1}  (1-p_i).
\end{equation}
The log-likelihood is then
\begin{equation}
    \ln \mathcal{L}_{\mathrm{A}}= \sum_i  (s_A(\Lambda_i)-1) \ln p_i +   \ln (1-p_i).
\end{equation}
Optimal values of $p_i$ are given by setting $\partial \ln \mathcal{L}_{\mathrm{A}}/\partial p_i =0 $:
\begin{equation}
    \frac{ s_A(\Lambda_i)-1 }{p_i} = \frac{1}{1-p_i},
\end{equation}
or 
\begin{equation}
   p_i = \frac{s_A(\Lambda_i)-1}{s_A(\Lambda_i)}, 
\end{equation}
which indeed is the survival probability factor in Eq.~\eqref{eq:drop3}, and leads to the total crossing probability $P_A(\lambda)$.
 This estimator for the survival probability for each path's extreme value $\Lambda_i$
turns out to be  related to the 
Kaplan-Meier
product-limit estimator with left truncation \cite{Kaplan1958,LyndenBell1971,Woodroofe1985}. In the next section we will make this connection stronger.

\subsection{Some algorithmic considerations}

The original WHAM approach in Ref.\cite{Rogal2010} is literately a histogram reweighting technique,  which   constructs a set of histograms based on the path ensemble data, and then combines these using the iterative algorithm. In the general case  $M\neq K$, i.e., the number of bins M in the histogram is not the same as the number of interface simulations. This is illustrated in Fig.\ref{fig:WHAMexample}, where the histograms contain many more bins than there are interfaces.
This choice also meant that the direct non-iterative solution  \ref{eq:titus} was   not available. Instead the  weights  $w_k$ from  the WHAM procedure, were used to define the weight for each path $w[x_{k,n}]$. 

While this histogram approach is valid, there are some subtleties involved. For instance, in  implementations\cite{Swenson2019a} one often applies a 'cutoff' for the histograms. If a path in an interface ensemble spuriously reaches an interface far way, this was found to influence the outcome of the WHAM significantly. As the resulting histogram statistics was considered untrustworthy, histograms were cut off at a certain threshold, usually something like 5-10\%.  In the Appendix \ref{app:A} we  examine this effect in more detail.

Later,  Du et al.\cite{Du2013} and Newton et al.~\cite{Newton2017a,Newton2017b}  introduced the  path numbers: the number of paths that have their maximum between two interfaces. This  effectively sets $M = K$. Iterative WHAM could still be applied to these path number 'histograms'. 
%\subsection{variance based method}
The introduction of 
van Erp's VBR %variance based reweighting
method\cite{vanErp2016} eliminated the need for iteration, by computing the crossing probabilities at each next interface directly from the previous ones. Thus, this approach only takes into account the path numbers.  
For both these $M = K$ approaches a cutoff is not needed.

%\subsection{MBAR}
The MBAR method for path sampling was introduced relatively late\cite{Breebaart2026combining}. This method has as an advantage that the MLE derivation is relatively straightforward, it is bin-less, and  does not assume specific variance properties. 
Moreover, it can be applied on the path level, giving a path weight directly to each path in the ensemble. 
The MBAR is in its general form also iterative but can be made non-iterative fairly easily, as we showed above. 
Remarkably, the  non-iterative way is not identical to the non-iterative VBR method, yet gives the same results. 
We note that MBAR also allows to combine sets of TIS based on different interface definitions\cite{Breebaart2026combining,Breebaart2026training}. In this case there is no non-iterative solution.

The SAR is a latest step in the evolution of the RPE approaches, and allows to combine TIS simulations by reordering trajectories based on the maximum lambda value. 
While it can be directly applied to the TIS ensembles, it is not limited to sampling TIS, but e.g. also using rejection free path sampling\cite{Lazzeri2025archive}, in which the shooting point itself acts as the interface.
The SAR method can be shown to be related to the  nonparametric maximum- likelihood estimator of product-limit form, a well known result in statistics but up to recently not much used in the enhanced sampling community\cite{Kaplan1958,LyndenBell1971,Woodroofe1985}.

\section{Synthesis of the approaches} 
\label{sec:synthesis}

In the previous sections we laid out the different routes to the RPE, as they were introduced in the literature, and indicated how they are connected.
However, several questions remain: Why does the reweighting work at all? Is there an underlying reason why all these routes amount to the same outcome? What exactly is the basis of the RPE? Why is it sufficient to record the   $\lambda_{max}$ for the reweighting? Why is it allowed to merge all data from separate biased ensembles?
In this section we are synthesizing our knowledge and give a fresh perspective on these questions.

\subsection{Conditioning on  $\lambda_{\rm max}$ allows factorization}
We start with a couple of definitions. As before, we have   $\PA[\XP]$ denoting the 
distribution of excursions, trajectories leaving A, and ending in A or B, normalized such that $\intp \PA[\XP] = 1 $
\label{eq:norm}. For each path we compute the maximum value of the CV or order parameter, $\lambda_{max}[\XP]$. 
The TIS interface ensembles are 
\begin{equation}
\PAk{k}[\XP] \;=\; \frac{1}{Z_k}\, {h}_k(\lambda_{max}[\XP])\, \PA[\XP],
\label{eq:pak}
\end{equation}
with $
h_k(\lambda) = \theta(\lambda-\lambda_k)$ and $Z_k = \intp \PA[\XP]\, h_k(\lmax[\XP])$. From the definition it is clear that $Z_k= P_A(\lambda_k|\lambda_0)$, i.e. the crossing probability at the interface $k$. 
Now, instead of constructing the RPE by reweighting, we first focus on the paths that have identical $\lmax=\Lambda$, by computing the density of paths
\begin{equation}
p(\Lambda) \;=\; \intp \PA[\XP]\,\delta(\lmax[\XP]-\Lambda),
\label{eq:marginal}
\end{equation}
where we used the Dirac-delta function to select just those paths that have their maximum at $\Lambda$.   This density is related to the crossing probability by simply integrating over $\Lambda$
\begin{align}
    P_A(\lambda|\lambda_0)= \int_\lambda^\infty d\Lambda p(\Lambda).
\end{align}
Now consider the pair ($\XP,\Lambda)$ and  apply Bayes' rule for their joint distribution 
\begin{equation}
\PA[\XP]\; p(\Lambda\,|\,\XP) = p(\Lambda)\; \PA[\XP\,|\,\Lambda].
\label{eq:bayes}
\end{equation}
Since  $\Lambda= \lmax[\XP]$ is a deterministic function of $\XP$, it follows that $p(\Lambda\,|\,\XP) \;=\; \delta(\Lambda-\lmax[\XP])$. This means that for the case that $\Lambda= \lmax[x]$, we obtain
\begin{equation}
%\PA[\XP] \;=\; p(\lmax[\XP]) \cdot \PA[\XP \,\big|\, \lmax[\XP]].
\PA[\XP] = p(\Lambda) \PA[\XP \,|\, \Lambda ].
\label{eq:factorization}
\end{equation}
This factorization is the basis for the reweighting. It clearly shows how each conditional path probability can be related to the full path ensemble $\PA[\XP]$ using a  factor  $p(\Lambda)$.  Indeed this has the same form as  Eq.~\ref{eq:GLseq}. 

\subsection{All ensembles along $\lmax$ are equivalent}
The factorization remains valid  for each biased ensemble $k$. To see this, first compute the marginal 
\begin{align}
p_k(\Lambda)
&= \intp \PAk{k}[\XP]\;\delta(\lmax[\XP]-\Lambda) \notag\\
&= \frac{1}{Z_k}\intp h_k(\lmax[\XP])\,\PA[\XP]\;\delta(\lmax[\XP]-\Lambda)\notag \\
&= \frac{1}{Z_k}\, h_k(\Lambda) \intp \PA[\XP]\;\delta(\lmax[\XP]-\Lambda)\notag\\
&= \frac{h_k(\Lambda)\; p(\Lambda)}{Z_k},
\label{eq:tiltedmarginal}
\end{align}
where  in the third line we used that, because of the delta function, $\Lambda = \lmax[\XP]$, so that the indicator $h_k$ can be taken outside of the integral. 
Using Bayes rule again, but now for $P_{A,k}[\XP]$, and following the same reasoning as above, inserting both Eq.\ref{eq:tiltedmarginal} and \ref{eq:pak}, we obtain for $\Lambda\ge\lambda_k$
\begin{equation}
\PAk{k}[\XP\,|\,\Lambda]
= \frac{\PAk{k}[\XP]\;p_k(\Lambda|\XP)}{p_k(\Lambda)}
= \PA[\XP\,|\,\Lambda]
%\quad (\Lambda\ge\lambda_k),
\label{eq:bayesinv}
\end{equation}
Remarkably, in the second equality the dependence on $k$ vanishes. This means that the distribution of paths within the $\Lambda$ slice is not altered by the bias, precisely what  TIS is supposed to do: sampling paths conditioned on an interface crossing without changing the relative path probability within a interface ensemble.
This independence means that for a certain $\Lambda$ the paths in all biased ensembles are samples of  the  conditional $\PP_A(\XP|\Lambda)$. Therefore, the data from different ensembles can be merged. 

Rearranging the 2nd equality yields for  $\Lambda= \lmax[\XP]$
\begin{align}
   \PAk{k}[\XP] = p_k(\lmax[\XP])\cdot \PA[\XP\,\big|\,\lmax[\XP]],
   \label{eq:factorizationbiasens}
\end{align}
Thus, up to single weighting factor, the biased ensembles are the same. This validates the assumption made in section \ref{sec:RPE}, e.g in Eq.~\ref{eq:biasscaling}.

\subsection{The weight is one-dimensional}
To compute this weight we can either just apply Eq.\ref{eq:pak}, or we can merge the obtained K  biased datasets where each path is drawn from the merged distribution  
\begin{equation}
q[\XP]
\;=\; \sum_{k=0}^{K-1} {N_k} \, \PAk{k}[\XP]
\;{=}\;
\Biggl[\, \sum_{k=0}^{K-1} {N_k}\, \frac{h_k(\lmax[\XP])}{Z_k} \Biggr]\, \PA[\XP].
\label{eq:mixture}
\end{equation}
To map the merged dataset back to the unbiased ensembles, each path $\XP$ receives  a weight 
\begin{equation}
\frac{\PA[\XP]}{q[\XP]}\;\propto\;  w[\XP] 
\;=\; \Biggl[\, \sum_{k=0}^{K-1} \frac{N_k %\mathcal{Z}
} {Z_k} {h_k(\lmax[\XP])}\Biggr]^{-1},
\label{eq:mbarweight}
\end{equation}
which is indeed the MBAR result, and also equivalent to the WHAM and VBR. 
Clearly, this weight purely only depends on $\Lambda= \lmax[\XP]$.

\subsection{The sufficiency of $\Lambda$}
Is $\Lambda$ a sufficient statistics to do the reweighting? To argue this  let us look at the likelihood maximization procedure itself. 
In practice,   $p(\Lambda)$ and  $Z_k$ are unknown, and need to be estimated, after which the path weights then follow automatically. 
Consider modeling the unknowns non-parametrically,
using as  data  the merged sets of  paths $\XP_{k,n}$ with maxima $\Lambda_{k,n}=\lmax[\XP_{k,n}]$.  The likelihood for  one sample is then, using the  factorization Eq.~\eqref{eq:factorizationbiasens} and  Eq.~\eqref{eq:tiltedmarginal}:
\begin{align}
\PAk{k}[\XP_{k,n}]
\;=\; p_k(\Lambda_{k,n})\;\mathcal{P}_A[\XP_{k,n}\,\big|\,\Lambda_{k,n}]\notag\\
\;=\; \frac{h_k(\Lambda_{k,n})\; p(\Lambda_{k,n})}{Z_k(p)}\;\mathcal{P}_A[\XP_{k,n}\,\big|\,\Lambda_{k,n}].
\label{eq:onesample}
\end{align}
The total likelihood follows from multiplying over the dataset:
\begin{align}
L% (p,\mathcal{C})
&= \prod_{k=0}^{K-1}\prod_{n=1}^{N_k} \PAk{k}[\XP_{k,n}]\notag\\
&= \underbrace{\prod_{k,n} \frac{h_k(\Lambda_{k,n})\; p(\Lambda_{k,n})}{Z_k(p)}}_{\displaystyle g(\{(k,\Lambda_{k,n})\};\,p)}
\;\times\;
\underbrace{\prod_{k,n} \mathcal{P}_A[\XP_{k,n}\,\big|\,\Lambda_{k,n}]}_{\displaystyle \text{free of } p}.
\label{eq:fisherneyman}
\end{align}
The first factor depends on the data only through the labeled maxima $(k,\Lambda_{k,n})$, whereas the second factor does not contain $p$ at all. 
Given the labeled maxima, the actual behavior of  the dynamical paths  is governed by $\mathcal{P}_A$ alone and contains no further information about $p$.  
The labeled maxima $\{(k,\Lambda_{k,n})\}$  are therefore a sufficient statistic for  $p$. Since the crossing probability,  path weights, and $Z_k$ 
are  functionals of $p$, no more information can be extracted about them from the paths than from the first factor, i.e., the maxima and  ensemble labels.

Maximizing the likelihood over $p$ involves only the first factor, which is the likelihood of the samples of $\Lambda$, each observed under a nested indicator bias. 
Thus,  every observation $\Lambda_{k,n}$ is a value drawn from $p$, \emph{left-truncated} at $\lambda_k$, i.e., we only observe it because it exceeded $\lambda_k$, while reactive paths are \emph{right-censored} at $\lambda_B$. The nonparametric maximum-likelihood estimator for truncated and censored data has a product-limit form, known as  the Kaplan-Meier estimator in its left-truncation variant
\cite{Kaplan1958,LyndenBell1971,Woodroofe1985}.  WHAM with one bin per interface, MBAR,  VBR, and  SAR are four different routes to this one estimator, which is why they  ultimately give the same answers.

\subsection{Transforming to hazard coordinates yields  SAR}
While sufficiency explains why every method must build on the labeled maxima, it does not yet explain why WHAM, MBAR,  VBR, and  SAR return the \emph{same} numbers. In  section III   we verified the equivalence pairwise.
To derive their equivalence to the product-limit estimator with left truncation,  we can apply a change into  \emph{hazard} variables.

We start by ordering the  observed maxima $\Lambda_{(1)}<\Lambda_{(2)}<\dots<\Lambda_{(N)}$ for the $N=\sum_k^{K-1} N_k$ paths. 
Next, we define the \emph{hazard}, $q_i$, as the probability that a path reaches a maximum value of $\lambda$ no greater than $\Lambda_i$, before returning to terminate at $\lambda_A < \Lambda_i$.
Thus, a path surviving up to $i-1$ and then dying at $i$ has a probability 
\begin{equation}
% q_i = \mathrm{Prob}(\Lambda = \Lambda_{(i)} \,\big|\, \Lambda \ge \Lambda_{(i)}),
% \qquad
p_i = q_i \prod_{j<i}(1-q_j),
\end{equation}
and the survival probability for a path to reach $\lambda$ is
\begin{equation}
S(\lambda) = \!\!\prod_{j:\,\Lambda_{(j)}<\lambda}\!\!(1-q_j).
\label{eq:survival}
\end{equation}

Consider now taking a path $n$, entering observation at its interface $e_n=\lambda_{k(n)}$ and dying at $\Lambda_n= \Lambda_{(i(n))}$. By Eq.~\eqref{eq:fisherneyman} this path contributes $p(\Lambda_n)/Z_{k(n)} = p(\Lambda_n)/S(e_n)$ to the  likelihood. Writing the log of  this factor  in the hazard coordinates gives:
\begin{align}
\ln\frac{p(\Lambda_n)}{S(e_n)}
&=\ln q_{i(n)} + \hspace{-0.2cm} \sum_{j<i(n)} \ln(1-q_j) - \hspace{-0.4cm}
\sum_{j:\,\Lambda_{(j)}<e_n} \ln(1-q_j), \notag
\\
% &=
% \ln q_{i(n)} \;+ \sum_{j:\; e_n \le \Lambda_{(j)} < \Lambda_n} \ln(1-q_j)\\
&\equiv
\ln q_{i(n)} \;+ \sum_{j\in W_n} \ln(1-q_j),
\label{eq:pathlogL}
\end{align}
where $W_n=\{j:\, e_n \le \Lambda_{(j)} < \Lambda_n\}$ is the path's \emph{observation window}, during which it survived the other paths $j$.  
Note how in the first sum of the first equality the paths that survive below $\Lambda_n$ split into those below $e_n$, which cancel against those in the last sum, and those in between $e_n$ and $\lambda_n$, which survive.
A path contributes $\ln q$ at its own death and $\ln(1-q)$ at every observed maxima inside its window. 
Note that a  reactive path contributes a factor $S(\lambda_B)/S(e_n)$   along its whole window and has no death term.

When  summing  Eq.~\eqref{eq:pathlogL} over all paths we can exchange the order of the double sum to yield \begin{equation}
\sum_n \sum_{j\in W_n} \ln(1-q_j)
\;=\; \sum_{j=1}^{M} \ln(1-q_j)\times \#\{n:\, j\in W_n\},\notag
\label{eq:swap}
\end{equation}
where the $\#\{.\}$ function counts the elements in the set.  Next we define, for each observed maximum $\Lambda_{(j)}$, the death and the risk set,
\begin{equation}
d_j = \#\{n:\, \Lambda_n = \Lambda_{(j)}\},
\qquad
r_j = \#\{n:\, e_n \le \Lambda_{(j)} \le \Lambda_n\},\notag
\label{eq:riskdef}
\end{equation}
Note that $\#\{n: j\in W_n\} = \#\{n:\, e_n\le\Lambda_{(j)}<\Lambda_n\} = r_j - d_j$ are the paths whose window contains  $\Lambda_{(j)}$, and  are exactly given by those at risk  minus those dying in that window. 
Hence, the likelihood for the entire set of paths is 
\begin{equation}
\ln L(q) \;=\; \sum_{i=1}^{N} [\, d_i \ln q_i \;+\; (r_i-d_i)\,\ln(1-q_i) \,]
\label{eq:binomial}
\end{equation}
The likelihood of the entire TIS dataset is thus a product of independent binomial experiments, one per observation. Reactive paths contribute survival factors only and simply enlarge the risk sets.

A simple likelihood estimation, setting the derivatives to zero,  gives 
\begin{equation}
\frac{\partial \ln L}{\partial q_i}
= \frac{d_i}{q_i} - \frac{r_i-d_i}{1-q_i} = 0,
\end{equation}
leading to the opimal value for $q$
\begin{equation}
\hat q_i = \frac{d_i}{r_i}.
\label{eq:qhat}
\end{equation}
The survival function $S(\lambda)$, equivalent to the crossing probability is then the  product-limit estimator 
\begin{align}
\hat S(\lambda)= P_A(\lambda|\lambda_0)= \prod_{\Lambda_{(i)}<\lambda}(1-d_i/r_i), 
\end{align} 
which is indeed equivalent to the SAR (see also Fig.~\ref{fig:dropmethod}). MBAR follows from a analog fashion, by optimizing the likelihood of Eq.~\ref{eq:fisherneyman}, as done in section \ref{sec:RPE}. 

\subsection{Final remarks}
To finish the above synthesis we make a few final observations. 
First, the non-iterative sweeps are possible because the map from hazards to survival values $p$ is \emph{triangular} since Eq.~\eqref{eq:survival} shows that $Z_k=S(\lambda_k)$ involves only hazards below $\lambda_k$. This means  the  equations in $Z_k$ can be solved forward in a single iteration. 

Second, the naive point-matching optimizes a \emph{pruned} likelihood, in which the risk set of every segment is reduced to the paths dying in that segment. 
Thus  this amounts to the  same model but with discarded data, and the estimator remains consistent but  precision and efficiency decreases. 

Third, histogram-based WHAM with more bins than interfaces solves the iterative equations for a \emph{misspecified} likelihood in which the cumulative crossing histogram registers
a single path as  counts in multiple bins below its maximum, double-counting it relative to Eq.~\eqref{eq:binomial}, such that the 
estimators demonstrably differ.  In the Appendix \ref{app:A} we show that 
this really makes a difference, and what the influence of  a histogram cutoff is.

\begin{table*}[t]
\centering
\caption{SAR  in terms of the hazard coordinate, applied to the three-interface example. Each bin has a risk set $r_j$ (entered, not yet dead), deaths $V_j$, survival after the bin, and the weight $w_j=\Delta\hat S_j/V_j=\hat S(\lambda_j)/r_j$ shared by all paths ending in the bin.}
\label{tab:ledger}
\begin{tabular}{lccccc}
\toprule
bin & new entrants & at risk $r_j$ & deaths $V_j$ & $\hat S$ after bin & weight per path \\
\midrule
$[\lambda_0,\lambda_1)$ & 1000 (ens.\ 0) & $1000$ & $500$ & $1\cdot\frac{500}{1000}=\frac12$ & $\frac{1-1/2}{500}=1/1000$ \\[3pt]
$[\lambda_1,\lambda_2)$ & 1000 (ens.\ 1) & $500+1000=1500$ & $1200$ & $\frac12\cdot\frac{300}{1500}=\frac1{10}$ & $\frac{1/2-1/10}{1200}=1/3000$ \\[3pt]
$[\lambda_2,\lambda_B)$ & 1000 (ens.\ 2) & $300+1000=1300$ & $1040$
  & $\frac1{10}\cdot\frac{260}{1300}=\frac1{50}$ & $\frac{1/10-1/50}{1040}=1/13000$ \\[3pt]
$\lambda_{\max}\ge\lambda_B$ (reactive) & 0 & $260$ & $260$ (censored)
  & --- & $\frac{1/50}{260}=1/13000$ \\
\bottomrule
\end{tabular}
\end{table*}

\section{Illustrative simple examples}

In this section we illustrate the different reweighting expressions by applying them to very simple pedagogical examples.  These examples are meant to clarify the inner workings of the reweighting.

\subsection{Single iteration VBR}
In the first  example we illustrate the non-iterative VBR method. 
To make the example concrete we borrow from the discussion of Ref.~\cite{Newtonthesis} using an extremely simple numerical example.
Consider a barrier crossing with four interfaces, $\{\lambda_0, \lambda_1, \lambda_2, \lambda_3\}$ 
and three ensembles $[0^+], [1^+],[2^+]$ with 1000 paths per ensemble: $N_0=N_1=N_2=1000$. 
 Further, we set $\lambda_0=\lambda_A$ and $\lambda_3=\lambda_B$. 
Instead of trying to obtain  the crossing probability  for each interface we now imagine that they are given to be  $P_A(\lambda_0|\lambda_0)=\ZZ_0/\ZZ =1$
, $P_A(\lambda_1|\lambda_0)=\ZZ_1/\ZZ=0.5$, $P_A(\lambda_2|\lambda_0)= \ZZ_2/ \ZZ=0.1$, and $P_A(\lambda_3|\lambda_0)= \ZZ_3/ \ZZ=0.02$. If we would perform a TIS simulation for this system and generate 1000 pathways per interface, we would get the following table, which  reports the expected total 
number of pathways $m_i(\lambda_j)$ (not to be confused with the path numbers $n_i(\lambda_j)$) per interface ensemble
$i$ (column) having a $\lambda_{\rm max} > \lambda_j$ for $j=0,1, 2, 3.$ (row).

\begin{center}
\begin{tabular}{c|ccc|c}
count & $[0^+]$ & $[1^+]$ & $[2^+]$ & $\sum_k m_k(\lambda_j)$\\
\hline
$\lambda_{\rm max} > \lambda_0$ & 1000 &   &  & 1000  \\
$\lambda_{\rm max} > \lambda_1$ & 500  & 1000 &  &1500  \\
$\lambda_{\rm max} > \lambda_2$ & 100   & 200  & 1000 &1300\\
$\lambda_{\rm max} > \lambda_3$ (reactive) & 20    & 40   & 200  & 260\\
%sum  & 1600   & 200  & 1000 \\
\end{tabular}
\end{center}
where the blank entries  denote the parts of the unnormalized histograms that do not contribute to the reweighting, i.e.  $j< i$. Note that this represents an idealized scenario in the sampling following \emph{exactly} the same  trend for each ensemble. 
Here, the overall crossing probability 
can even be correctly obtained from the $[0^+]$ ensemble as: $P_A(\lambda_3|\lambda_0)=
m_0(\lambda_3)/m_0(\lambda_0)=
20/1000=0.02$. 
If we would use the point-matching method Eq.~\ref{eq:pointmatching}:
\begin{align}
P_A(\lambda_3|\lambda_0) &=
\frac{m_0(\lambda_1)}{m_0(\lambda_0)} 
\frac{m_1(\lambda_2)}{m_1(\lambda_1)}
\frac{m_2(\lambda_3)}{m_2(\lambda_2)}\notag\\ &=
\frac{500}{1000} \frac{200}{1000} \frac{200}{1000} = 0.02
\notag
\end{align} 
we also obtain the same result.

Now, let us apply the VBR Eq.~\ref{eq:titus}, in which $P(\lambda_3 | \lambda_0)$ is computed in a single iteration over three steps, with $\lambda$ successively increased from $\lambda_1$ to $\lambda_2$ and subsequently to $\lambda_3$:\\
%\small
%\begin{widetext}
\begin{align*}
{\bf Step 1:}  &
\lambda=\lambda_1.  \notag \\
& %K(\lambda_1)
k=0  \notag \\
& P_A(\lambda_1|\lambda_0) =
\frac{m_0(\lambda_1)}{N_0 [P_A(\lambda_0|\lambda_0)]^{-1}} \notag \\& =
\frac{m_0(\lambda_1)}{N_0}=\frac{500}{1000}=0.5
\notag \\
{\bf Step 2: } &
%Eq.~\ref{eq:Titus} with 
\lambda=\lambda_2.\notag \\
& %K(\lambda_2)
k=1\notag \\
& P_A(\lambda_2|\lambda_0) =
\frac{m_0(\lambda_2) + m_0(\lambda_2)}{N_0 [P_A(\lambda_0|\lambda_0)
]^{-1}+
N_1 [P_A(\lambda_1|\lambda_0)
]^{-1}
} \\ & =
\frac{100+200}{1000+1000/0.5}=\frac{300}{3000}=0.1
\notag %\\
\end{align*}
\begin{align*}
{\bf Step 3: } &
\lambda=\lambda_3.\notag \\
& %K(\lambda_3)
k=2\notag \\
& P_A(\lambda_3|\lambda_0)= 
\frac{m_0(\lambda_3) + m_1(\lambda_3)+m_2(\lambda_3)}{\frac{N_0} {P_A(\lambda_0|\lambda_0)}+
\frac{N_1} {P_A(\lambda_1|\lambda_0)} +
\frac{N_2} {P_A(\lambda_2|\lambda_0)}
} \\ &= 
\frac{20+40+200}{1000+1000/0.5+1000/0.1}=\frac{260}{13000}=0.02\notag \\
\end{align*}
%\end{widetext}
% \begin{align}
\normalsize

This demonstrates that, in this idealized scenario, all three approaches, and by extension  WHAM/MBAR, yield the same result.

\subsection{Interface crossings vs path numbers}
\label{sec:example1}

Next we compare two  seemingly different expressions for the path weights. The first expression is  the WHAM Eq~\ref{eq:whamweights}-\ref{eq:whamselfconsist}, or equivalently  Eqs.~\ref{eq:MBARWHAMeq},~\ref{eq:pathnumbers} or \ref{eq:varbasedreweighting}, which all count the path crossings of the interfaces. The second expression is given by  Eq. \ref{eq:MBARWHAMeq4}, which focuses on the path numbers $n_i(\lambda_j)$, i.e., how many paths reach their maximum in $\D_j$.
%by  the 
%maximum reached $\lambda$  for the paths.

We consider again the same example as in the previous section.
The WHAM weights in  Eq.~\ref{eq:whamweights}  become  $w_0 = 1/1000$, $w_1= 1/3000$, $w_2=1/13000$ and $w_3=w_2$ since $N_3=0$ as no new trajectories are introduced at $\lambda_3$. Applying then the reweighting according to Eq.~\ref{eq:wham} or  Eq.~\ref{eq:varbasedreweighting}, gives us back the correct crossing probability weights, as shown in the previous section.

However, instead of the crossings we could also look at the maximum crossed interface for each path to obtain the path number matrix
\begin{center}
\begin{tabular}{l|ccc|c}
count & $[0^+]$ & $[1^+]$ & $[2^+]$ & $V_j=\sum_k n_k(\lambda_j)$\\
\hline
$\lambda_0 < \lambda_{\rm max} < \lambda_1$ & 500 &   &  & 500  \\
$\lambda_1 <\lambda_{\rm max} < \lambda_2$ & 400  & 800 &  &1200  \\
$\lambda_2  < \lambda_{\rm max} < \lambda_3 $ & 80   & 160  & 800 &1040\\
$\lambda_{\rm max} > \lambda_3$ & 20    & 40   & 200  & 260 \\
%sum  & 1600   & 200  & 1000 \\
\end{tabular}
\end{center}
which clearly differs from the  table in the previous section.
Yet, now applying the single iteration MBAR Eq.~\ref{eq:MBARWHAMeq6} also leads to the unbiased path number distribution, specifically
$V_0 w_0 = 0.5$,  $V_1 w_1 = 0.4$, $V_2 w_2 = 0.08$ and $V_3 w_3 = 0.02$, which by the  summation in Eq.~\ref{eq:MBARWHAMeq5}, then also gives back the correct crossing probabilities, respectively $Z_0=1, Z_1= 0.5, Z_2=0.1,Z_2=0.8$.

Thus,  remarkably, reweighting using interface crossing  numbers and using the path numbers based on the maxima leads to exactly the same results.

\begin{table*}[t]
    \centering
    \caption{MBAR/SAR analysis of synthetic data for linear barrier crossing.\label{tab:syndata}}
    \begin{tabular}{c|cccccccccc}
interface $i$ & 0 & 1 & 2 & 3 & 4 & 5 & 6 & 7 & 8 & 9 % & 10
\\
\hline
$N_i$  & 50 & 50 & 50 & 50 & 50 & 50 & 0 & 0 & 0 & 0 \\
$V_i$ &  30 & 38 & 51 & 52 & 48 & 50 & 0 & 0 & 31 & 0 \\
$S_i$ & 50 & 100 & 150 & 200 & 250 & 300 & 300 & 300 & 300 & 300 \\
$F_i$ &30 & 68 & 119 & 171 & 219 & 269 & 269 & 269 & 300 & 300 \\
 $S_i-F_i+V_i$ & 50 & 70 & 82 & 81 & 79 & 81 & 31 & 31 & 31 & 0 \\
$Z_i$ & 1.00 & 0.400 & 0.183 & 0.0691 & 0.0247 & 0.00971 & 0.00372 & 0.00372 & 0.00372 & 0 \\
$w_i$& 0.0200 & 0.00571 & 0.00223 & 0.000853 & 0.000313 & 0.000120 & 0.000120 & 0.000120 &
   0.000120 & 0 \\
\end{tabular}
\end{table*}

\begin{figure*}[t]
    \centering
    \includegraphics[width=0.8\linewidth]{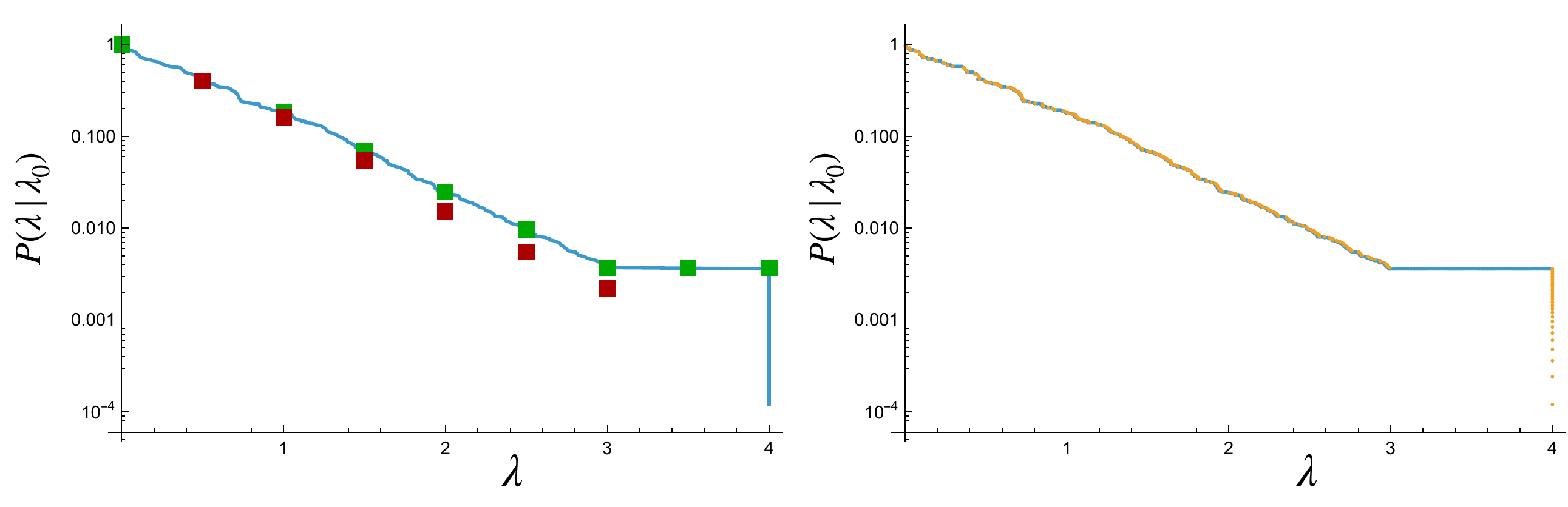}
    \caption{Reweighted crossing probability for simple linear potential. Left: red squares: point-wise matching, green curve: non-iterative MBAR, blue curve: SAR.  
    Right:  comparison between SAR, and  RPE based probability. Note that all approaches give identical results, except for the  the point-wise matching, which gives substantially lower estimates.}
    \label{fig:rpelinpot}
\end{figure*}

\subsection{SAR in terms of the hazard coordinate  }

Let us now apply the SAR along the hazard coordinate. The three ensembles at $\lambda_0<\lambda_1<\lambda_2$ contribute again $N_k=1000$ paths each.  Classifying  by  the bin of the maximum reached along $\lambda$, ensemble $0$ has $500$ paths dying in $[\lambda_0,\lambda_1)$, $400$ in $[\lambda_1,\lambda_2)$, $80$ in $[\lambda_2,\lambda_B)$ and $20$ reaching $\lambda_B$. 
Next, ensemble $1$ has $800$ dying in $[\lambda_1,\lambda_2)$, $160$ in $[\lambda_2,\lambda_B)$ and $40$ reaching $\lambda_B$.
Finally, ensemble $2$ has $800$ dying in $[\lambda_2,\lambda_B)$ and $200$ reaching $\lambda_B$.
% All $1000$ paths of ensemble $2$ reach $\lambda_2$. 
We now walk along the $\lambda$ axis once, keeping track  of the entries, risk sets, and deaths  in Table~\ref{tab:ledger}.

In the first bin only ensemble $0$ has entered, so $r_0=1000$. Then five hundred paths die, and the survival curve drops to $\hat S(\lambda_1)=\frac{1}{2}$. At $\lambda_1$, ensemble $1$ enters so that  the adjusted risk set becomes the $500$ survivors of ensemble $0$ plus the $1000$ newcomers. Of these $1500$, $1200$ die before $\lambda_2$, so $\hat S(\lambda_2)= \tfrac12\cdot\tfrac{300}{1500}=\tfrac1{10}$. At $\lambda_2$, ensemble $2$ enters, and the risk set becomes $300$ survivors plus the $1000$ newcomers. Of these $1300$, $1040$ die before $\lambda_B$, so $\hat S(\lambda_B)= \tfrac1{10}\cdot\tfrac{260}{1300}=\tfrac1{50}$. The remaining $260$ paths are reactive. They are censored at $\lambda_B$ rather than dying, and share the remaining survival $\hat S(\lambda_B)=\tfrac{1}{50}$. 
The per-path weights are thus $1/1000$, $1/3000$, $1/13000$ and $1/13000$---exactly the WHAM/MBAR weights $w_0, w_1, w_2$ and $w_3=w_2$ that we obtained in the  example in Sec.~\ref{sec:example1}. 
As a check, the total probability is conserved: $500\cdot\tfrac{1}{1000} + 1200\cdot\tfrac{1}{3000} + 1040\cdot\tfrac{1}{13000} + 260\cdot\tfrac{1}{13000} = \tfrac12+\tfrac25+\tfrac2{25}+\tfrac{1}{50} = 1$. 

Table~\ref{tab:ledger} shows how each bin contributes one binomial factor of Eq.~\eqref{eq:binomial}, where out of $r_j$ paths at risk, $d_j=V_j$ die. The reactive paths contribute no death term, but only enlarge the risk set of the bins they pass through. These independent binomials are maximized bin by bin, without any iteration. 
This illustrates that the SAR indeed requires no iteration, no self-consistency, and even no histograms.

\subsection{ MBAR and SAR for a linear potential}
In a slightly more realistic example with non-ideal data we examine  a linear barrier  described by a potential $v(x)= 2 x$, with a transition state located at $x=3$. After this value of $x$ the barrier is crossed and trajectories go on to B defined at $x=4$.  We define 9 interfaces at $\lambda=0.5 i$, at  values $\lambda =\{0., 0.5, 1., 1.5, 2., 2.5, 3., 3.5, 4., 4.5\}$.
For each interface we synthetically create a set of $N_i=50$ shots on this barrier, which reach  a maximum $\lambda$ value given by an exponential distribution  $\exp{(-v(x))}$.  If a path crosses the transition state value $x=3$ this path is considered to continue  to reach the final state, which we identify by $x=4$.
The relevant data is stored in the $V_i$, the number of paths that reach $i$ but not $i+1$. Using the MBAR non-iterative Eq.~\ref{eq:MBARWHAMeq6} and the SAR Eq.~\ref{eq:drop1} we compute $S_i,Fi$, giving rise to $Z_i$ and $w_i$., The results are given in Table~\ref{tab:syndata} and plotted in Fig.~\ref{fig:rpelinpot}. The MBAR results agree exactly with those of the SAR.  In the right panel of Fig.~\ref{fig:rpelinpot} we compare the SAR with the RPE-based crossing probability Eq.~\ref{eq:RPEcrossprob}, which indeed are also identical, as expected.  
In addition, we  show the results for the naive point matching Eq.~\ref{eq:pointmatching}, which clearly underestimates the final crossing probability due to the fact that data needed to be discarded.

\section{Discussion and conclusion}

In this review we discussed apart from the standard point-wise matching,  four different ways of combining TIS trajectory data to obtain the reweighted path ensemble: WHAM, MBAR, the VBR, and the SAR. We find that all 4 ways lead to the same equations, and are in fact  leading to the same numerical results. 
We showed that the traditional  WHAM and MBAR iterative solutions can be replaced with a single  non-iterative sweep.

In our comparison, we also found a, previously unrecognised, inconsistency in the reweighting when using WHAM with histograms with a number of bins larger than the number of interface ensembles. The fact that the  histogram has more than one entry in the crossing probability, has  spurious effects for data with an outlier, which led in the past to the use of a cutoff. We showed (in the Appendix) that indeed for WHAM a cutoff leads to improved results, but this cutoff is not necessary when using just one entry per interface, such as done in MBAR, the VBR  or the SAR methods.   Therefore we argue that the WHAM based approach using histograms is possibly the least trustworthy, and we recommend one of the other three  non-iterative approaches.

In addition, we established the connection between the SAR and existing methodologies in the statistics and astronomy literature. This led in turn to a  synthesis of the concepts entering all reweighting methods.
We can conclude therefore that the reweighting methods developed in the path sampling community over the past two decades are all different manifestations of statistically optimal methods, i.e. they are using the trajectory data in the best possible way. 
Therefore, the foundations of the different reweighting methods lie 
in the censored data non-parametric maximum likelihood  estimator. 
This connection clarifies the reweighting approaches  from a theoretical point of view. 
This means not only that the methodologies are build on solid foundation, but also that one can build   free on all the techniques that have been developed in that field, such as the estimator of the variance, conjugate prior for obtaining closed solutions of Bayesian posterior,  etc.~\cite{Kaplan1958,Woodroofe1985}

From an operational point of view the choice about which one to use is purely of algorithmic convenience. Iterative methods are often not needed, and simpler sequential methods can be used.

Finally, we stress that  in our treatment we never used the Markov property, detailed balance, reversibility, equilibrium, or any property of the dynamics itself. Instead  the argument is rooted in probability theory  and  could be in principle  applied  to underdamped, non-Markovian, or driven dynamics, as long as a ensemble of trajectories  undergoing the   excursions is sufficiently sampled. 
In addition, the analysis in principle 
%is in principle not limited to TIS, but 
could  apply to other path sampling algorithms such as Forward Flux Sampling\cite{Allen2006} or Adaptive Multilevel Sampling\cite{Cerou2011}.

\section*{acknowledgments}

T.v.E and P.G.B. acknowledge the Holland Research School for Molecular Chemistry (HRSMC) for a visiting scholarship that made part of this work possible. 
G.L. and R.C. acknowledge the support of the Goethe University
Frankfurt, the Frankfurt Institute for Advanced Studies,  and the LOEWE
Center for Multiscale Modelling in Life Sciences of the state of Hesse.

\section*{AI disclosure}
In  preparing this manuscript the authors tasked Claude Opus 4.8  (Anthropic)  to  find any (unexpected) connections between  the different reweighting methods and the existing statistics literature,  leading to the synthesis  described in Sec.~\ref{sec:synthesis}.  The output was reviewed and verified by the authors, who take full responsibility for its accuracy and take full responsibility for the content of the publication. 
No generative AI tools were used for the manuscript text.

\appendix

\begin{figure*}[t]
    \centering
    \includegraphics[width=0.75\linewidth]{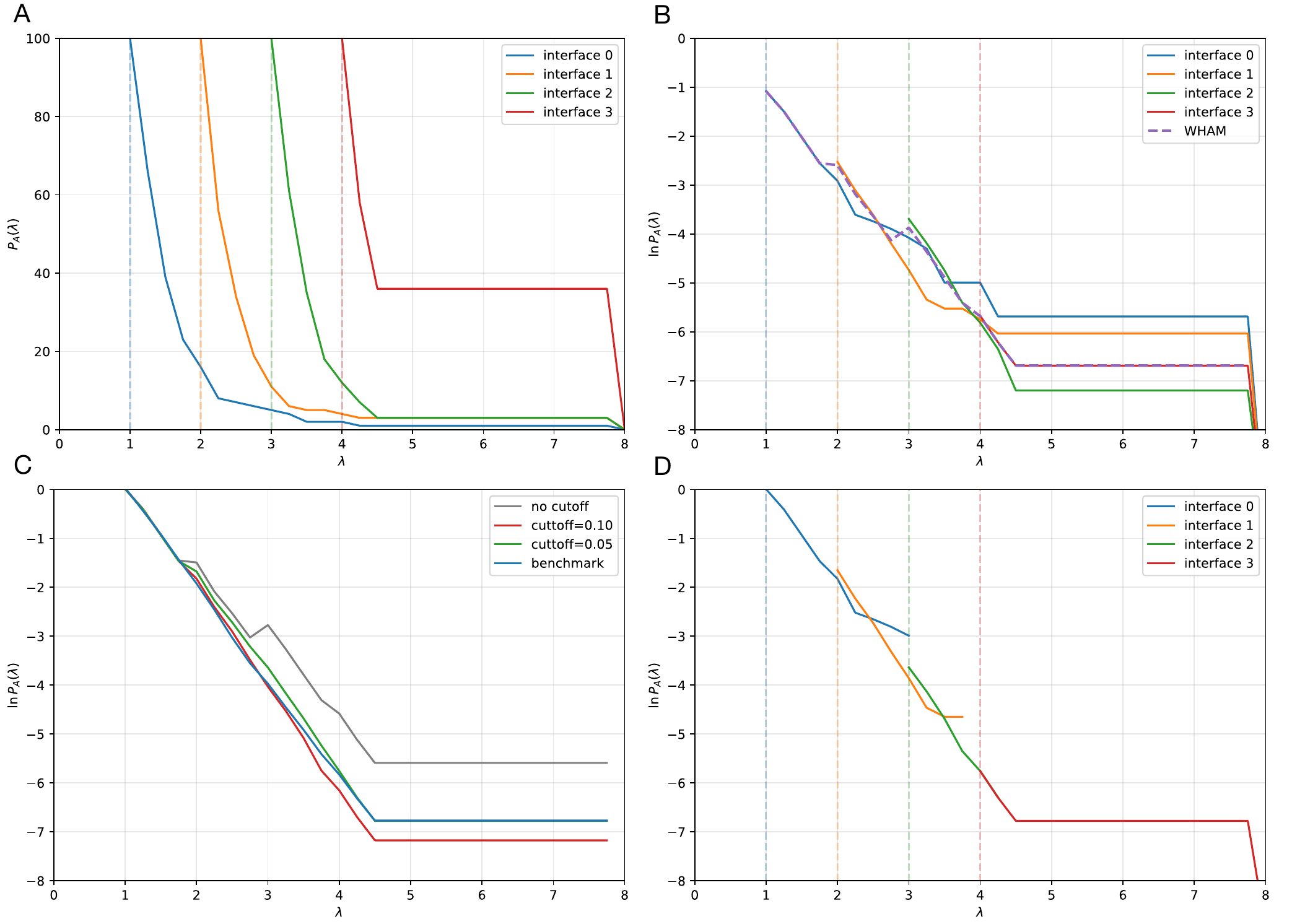}    
    \caption{A) Raw crossing probabilities obtained from synthetic data. B) Individual (unnormalised) crossing histograms after WHAM reweighting. The WHAMmed total histogram is in purple. C)  Effect of cutoff on the crossing probability. "no cutoff" data (gray) is the same is a the purple curve in panel B. The red and green curves are the WHAM results with a cutoff applied. The benchmark is based on more data. d) Individual WHAM histograms  result after application of the 0.05 cutoff.} 
    \label{fig:crosprob}
\end{figure*}

\begin{table*}[t]
\centering
\caption{All reweighting schemes unified in one set of quantities, for the three-ensemble example: three path ensembles of $1000$ paths each with $P_A(\lambda_1|\lambda_0)=0.5$, $P_A(\lambda_2|\lambda_0)=0.1$, $P_A(\lambda_3|\lambda_0)=0.02$. Column $j$ refers to interface $\lambda_j$ for crossing quantities and to the bin $[\lambda_j,\lambda_{j+1})$ for bin quantities; $j=3$ is $\lambda_B$, i.e.\ the reactive paths. All definitions follow from the path numbers $n_k(\lambda_j)$; 
%the 
%last text
third column:
%lists how
%the symbol under which 
how each quantity appears in the individual schemes (PM: point matching, PL: product-limit).}
\label{tab:A_methods_comparison}
\small
\renewcommand{\arraystretch}{1.2}
\setlength{\tabcolsep}{4pt}
%\begin{tabular}{>{\raggedright\arraybackslash}p{3.3cm} >{\raggedright\arraybackslash}p{3.6cm} >{\raggedright\arraybackslash}p{4.5cm} cccc}
 \begin{tabular}{p{3.3cm} p{3.6cm} p{4.5cm} cccc}
\toprule
Quantity & Definition & Appears as & $j=0$ & $j=1$ & $j=2$ & $j=3$ \\
\midrule
\multicolumn{3}{l}{\textit{Path numbers $n_k(\lambda_j)$, paths of $[k^+]$ with $\lambda_{\max}\in[\lambda_j,\lambda_{j+1})$}}\\
$n_0(\lambda_j)$ & & & 500 & 400 & 80 & 20 \\
$n_1(\lambda_j)$ & & & -- & 800 & 160 & 40 \\
$n_2(\lambda_j)$ & & & -- & -- & 800 & 200 \\
\midrule
\multicolumn{3}{l}{\textit{Crossing counts $m_k(\lambda_j)=\sum_{i\ge j}n_k(\lambda_i)$, paths of $[k^+]$ with $\lambda_{\max}\ge\lambda_j$}}\\
$m_0(\lambda_j)$ & & WHAM: $m_{A,0}(\lambda_j)$ & 1000 & 500 & 100 & 20 \\
$m_1(\lambda_j)$ & & WHAM: $m_{A,1}(\lambda_j)$ & -- & 1000 & 200 & 40 \\
$m_2(\lambda_j)$ & & WHAM: $m_{A,2}(\lambda_j)$ & -- & -- & 1000 & 200 \\
\midrule
\multicolumn{3}{l}{\textit{Path counts}}\\
$N_j$; \footnotesize sampled in $[j^+]$ & $\sum_{i\ge j}n_j(\lambda_i)$ & all: $N_j$;\ PL: new entrants & 1000 & 1000 & 1000 & 0 \\
$V_j$; \footnotesize ending in bin $j$ & $\sum_{k\le j}n_k(\lambda_j)$ & MBAR: $|D_j|$;\ SAR: $V_j$;\ PL: deaths $d_j$ & 500 & 1200 & 1040 & 260 \\
$S_j$; \footnotesize entered up to $\lambda_j$ & $\sum_{k\le j}N_k$ & SAR: $S_j$ & 1000 & 2000 & 3000 & 3000 \\
$F_j$; \footnotesize ended up to bin $j$ & $\sum_{k\le j}V_k$ & SAR: $F_j$ & 500 & 1700 & 2740 & 3000 \\
$r_j-N_j$; \footnotesize carried in & $\sum_{k<j}\sum_{i\ge j}n_k(\lambda_i)$ & PL: survivors & 0 & 500 & 300 & 260 \\
$r_j$; \footnotesize at risk at $\lambda_j$ & $\sum_{k\le j}\sum_{i\ge j}n_k(\lambda_i)$\newline $=S_j-F_j+V_j$ & WHAM: $m_A(\lambda_j)$;\ SAR: $s_A(\lambda_j)$;\ PL: $r_j$ & 1000 & 1500 & 1300 & 260 \\
\midrule
\multicolumn{3}{l}{\textit{Ratios and weights}}\\
$\rho_j$; \footnotesize local crossing ratio & $\sum_{i\ge j}n_{j-1}(\lambda_i)/N_{j-1}$ & PM: $P_A(\lambda_j|\lambda_{j-1})$ & -- & 0.5 & 0.2 & 0.2 \\
$q_j$; \footnotesize hazard in bin $j$ & $V_j/r_j$ & PL: $\hat q_j$;\ SAR: $1-\text{drop}$;\ MBAR: $w_jV_j/\mathcal Z_j$ & 0.5 & 0.8 & 0.8 & -- \\
$w_j$; \footnotesize weight per path & $1/w_j=\sum_{k\le j}N_k/P_A(\lambda_k|\lambda_0)$ & WHAM: $w_k$;\ MBAR: $w_{j,n}$;\ PL: $\hat S(\lambda_j)/r_j$;\ RPE: $w[\mathbf x]$ & 1/1000 & 1/3000 & 1/13000 & 1/13000 \\
$w_jV_j$; \footnotesize bin probability & & MBAR: $\mathcal Z_j-\mathcal Z_{j+1}$;\ path numbers: $\tilde n(\lambda_j)$ & 0.5 & 0.4 & 0.08 & 0.02 \\
\midrule
\multicolumn{3}{l}{\textit{Result $P_A(\lambda_j|\lambda_0)$}}\\
$[0^+]$ only & $\sum_{i\ge j}n_0(\lambda_i)/N_0$ & & 1 & 0.5 & 0.1 & 0.02 \\
point matching & $\prod_{i=1}^{j}\rho_i$ & PM: $\prod P_A(\lambda_{i+1}|\lambda_i)$ & 1 & 0.5 & 0.1 & 0.02 \\
pooled schemes & $\prod_{i<j}(1-q_i)=\sum_{i\ge j}w_iV_i$ & 
WHAM: $w_j\,m_A(\lambda_j)$;\newline 
MBAR: $\mathcal Z_j$;\newline 
VBR: $\tfrac{\sum_{i<j}m_i(\lambda_j)}{\sum_{k<j}N_k\,[P_A(\lambda_k|\lambda_0)]^{-1}}$;\newline SAR: $\prod_{i<j}[1-V_i/s_A(\lambda_i)]$;\newline 
PL: $\hat S(\lambda_j)$;\newline 
All RPE: $\sum_{\mathbf x}w[\mathbf x]\,\theta(\lambda_{\max}[\mathbf x]-\lambda_j)$ 
& 1 & 0.5 & 0.1 & 0.02 \\
\bottomrule
\end{tabular}
\end{table*}

\section{The influence of a cutoff on WHAM}
\label{app:A}
In this appendix  we investigate how WHAM responds to imposing a cutoff on the crossing probability histogram, in particular when the number bins in the histograms does not correspond to the number of interfaces.

We start with a linear barrier  described by $\beta v(x)= 2 x$, with a transition state located at $x=4.5$. We define four interfaces on $\lambda=x$, at integer values $\lambda =1,2,3,4$.
For each interface we  create a synthetical set of $n$ shots on this barrier, which reach until a maximum lambda given by the exponential distribution  $\exp(-\beta v(x))$.  If a path crosses the transition state value the path is considered to continue  to reach the final state, which we identify by $x> 8$.
To make the data set  sensitive to (sampling) errors we added to all interface sets a reactive path. 

For $n=100$  this leads to the  set of crossing probabilities shown in Fig.~\ref{fig:crosprob}a.
Superficially these  histograms   look fine. When applying the WHAM algorithm to this data we obtain the reweighted crossing probabilities, shown in   Fig.~\ref{fig:crosprob}b. 
%shows the logarithm of this crossing probability. 
In this plot the purple  dashed line is de  WHAMmed histogram, while the other curves correspond to the reweighted histograms from Fig.~\ref{fig:crosprob}a.
Note that the histograms do not completely line up due the additional error caused by the extra reactive  path per set. This has the effect that  in the lower interface sets the crossing probability for the higher interfaces are overestimated, i.e. the histograms are too shallow. To compensate this, the next interface needs to start a bit higher. For instance the blue curve does not decrease sufficiently steep.
%down steep enough. 
The orange histogram therefore starts a bit higher when optimally matching the histograms. This causes a bit (small) jump in the overall (purple) histogram. 

This is clearly unwanted, and is caused by the lower interface  paths that are reactive. Since the error in the histogram value itself is large due to the sparse data, this gives an additional   error when matching. 
The solution adopted in the literature %in  the early days 
was to cut the data at a specific threshold, above which the data was considered trustworthy.
In  Fig.~\ref{fig:crosprob}c we plot the effect of applying the cutoff.
Here, the gray curve is the same curve as the purple line in Fig.~\ref{fig:crosprob}b. It clearly overestimates the crossing probability  by a factor  3 compared to the benchmark (blue).
When applying the cutoff, the WHAM results converges to the benchmark.
The WHAMmed histograms for the 0.05 cutoff in Fig.~\ref{fig:crosprob}d look much more accurate, and coincide with the benchmark. 
Thus the truncated data set indeed seems to fit better than the uncut data.

Why is there such a difference? First, notice that when not applying the cutoff, the WHAM tries to align all histograms by an equal amount. In Fig.~\ref{fig:crosprob}b for instance the green histogram initially aligns well with the blue, while it actually should have aligned with the orange curve. This is likely caused by the fact that   all histogram points are treated a equally trustworthy in the method. Second, there is the compensation effect. If the slope of the (log) curves differ, the WHAM compensates by trying to matching somewhere in the middle, so that the  curves jump at the the interface value.
When the data is good, this is not a problem. However, when the slopes begin to differ, a more careful approach is needed. 

Finally, there is a obvious difference  between the WHAM\cite{Rogal2010} as we discuss it here and the MBAR method\cite{Breebaart2026combining} or the VBR method~\cite{vanErp2016}: namely, the treatment of the trajectory data is in the form of histograms in the case of WHAM and in the form of crossing numbers per interface in the case of MBAR/VBR method.  
The two methods coincide when the histograms are presented as as a single  bin per interface, but can differ when there are more bins between interfaces. 
In particular   when  histograms extend beyond the last interface (i.e. transition state), the histograms from the early interfaces can be untrustworthy, yet are treated with the same level of confidence as the later interface histograms. 
In the next section we examine how such outliers behave for  the correct WHAM-based expressions. 

\section{Influence of outliers for a single bin per interface}

Here we examine the influence of an outlier using the same very simple setup as used in section \ref{sec:example1}. 
Consider a case with a much lower crossing probability of $10^{-6}$, but containing a single fully reactive outlier in the $[0^+]$ ensemble. Such an event may arise from initialization memory effects and insufficient discarding of initial trajectories in the analysis. 

This could lead to the result table below:
%\begin{table}[h!]
\begin{center}
\begin{tabular}{c|ccc}
count & $[0^+]$ & $[1^+]$ & $[2^+]$ \\
\hline
$\lambda_{\rm max} > \lambda_0$ & 1000 & 1000 & 1000 \\
$\lambda_{\rm max} > \lambda_1$ & 10  & 1000 & 1000 \\
$\lambda_{\rm max} > \lambda_2$ & 1   & 10  & 1000 \\
$\lambda_{\rm max} > \lambda_3$ & 1    & 0   & 10  \\
\end{tabular}
%\end{table}
\end{center}

Here, the overall crossing probability 
obtained from the $[0^+]$ ensemble will give a huge overestimation:

\[ P_A(\lambda_3|\lambda_0)=
m_0(\lambda_3)/N_0=
1/1000=0.001 \gg 10^{-6}\]

Without outliers, one would expect the $[0^+]$ ensemble alone to yield zero when the overall crossing probability is very small and only 1000 trajectories are used, i.e., for a rare event typically treated with TIS methods. However, in this scenario, which may result from improper initialization, the crossing probability is vastly overestimated 
when based solely on the $[0^+]$ ensemble.

The point-match method is not affected by the outlier and yields the correct result:
\begin{align}
P_A(\lambda_3|\lambda_0)=
\frac{m_0(\lambda_1)}{N_0} 
\frac{m_0(\lambda_2)}{N_1}
\frac{m_0(\lambda_3)}{N_2} \notag \\=
\frac{10}{1000} \frac{10}{1000} \frac{10}{1000} = 10^{-6}    \notag 
\end{align}

Now, let us examine the effect of the outlier in the VBR expression Eq.~\ref{eq:titus} using the single iteration:

\noindent
\begin{align}
{\bf Step 1: } &
\lambda=\lambda_1. \notag\\
& %K(\lambda_1)
k=0 \notag\\
&P_A(\lambda_1|\lambda_0)=
\frac{m_0(\lambda_1)}{N_0 [P_A(\lambda_0|\lambda_0)]^{-1}}=
\frac{m_0(\lambda_1)}{N_0} \notag\\ &=\frac{10}{1000}=0.01 \notag
\\
{\bf Step 2: } &
%Eq.~\ref{eq:Titus} with 
\lambda=\lambda_2 \notag\\
&%K(\lambda_2)
k=1\notag\\
&P_A(\lambda_2|\lambda_0)=
\frac{
m_0(\lambda_3) + m_1(\lambda_3)}
%n_0[2] + n_1[2]}
{N_0 [P_A(\lambda_0|\lambda_0)
]^{-1}+
N_1 [P_A(\lambda_1|\lambda_0)
]^{-1}
} \notag\\ &=
\frac{1+10}{1000+1000/0.01}=\frac{11}{101000}=
1.0891089\cdot 10^{-4}
\notag %\\
\end{align}
\begin{align}
{\bf Step 3: } &
\lambda=\lambda_3.\notag\\
& %K(\lambda_3)
k=2\notag\\
& P_A(\lambda_3|\lambda_0)=
\frac{
m_0(\lambda_3) + m_1(\lambda_3)+m_2(\lambda_3)}
%n_0[3] + n_1[3]+n_2[3]}
{\frac{N_0} {P_A(\lambda_0|\lambda_0)}+
\frac{N_1}{P_A(\lambda_1|\lambda_0)}+
\frac{N_2} {P_A(\lambda_2|\lambda_0)}}
\notag\\
&=
\frac{1+0+10}{1000+1000/0.01+1000/1.0891089\cdot 10^{-4}} \notag\\
&=\frac{11}{9282818}=
1.185\cdot 10^{-6}
\end{align}

We observe that, in %the WHAM approach, using
the  VBR single iteration 
approach
based on Eq.~\ref{eq:titus}, 
the outlier does influence the final result, as it should when there is no clear justification to classify a trajectory as an outlier. However, the overall impact is modest: it increases the final estimate by only about 18\%. This stands in stark contrast to the result based solely on the $[0^+]$ ensemble, which was over three orders of magnitude too high.

\section{Example overview}

Table~\ref{tab:A_methods_comparison} summarizes all quantities for the reweighting schemes for the
%in terms of a
%single set of quantities for te three-
3-ensemble example of Sec.\ref{sec:example1}.

\section{Symbols}
   Tables \ref{tab:defs} and \ref{tab:defs2} in  list the  symbols and functions used in  the main text, together with a short explanation.
   
\begin{table*}[h!]
\centering
\caption{Explanation of used symbols} 
%Trajectories, states, and underlying dynamics.}
\label{tab:defs}
\begin{tabular}{ll}
\toprule
Symbol & Description \\
\midrule
&{Trajectories, states, and underlying dynamics.}\\
\midrule
$\XP$, & A trajectory (path), $\XP=\{x_0,x_1,\dots,x_L\}$ \\
$x_i$ & Phase-space configuration at frame $i$ (position, or position + momenta) \\
$L$ & Number of integration steps in a path \\
$\Delta t$ & Time step between frames \\
$\mathcal{T}=L\Delta t$ & Total duration of a path \\
$p(x_i\rightarrow x_{i+1})$ & Short-time Markovian transition probability (the unbiased dynamics) \\
$\rho(x)=e^{-\beta U(x)}/Z$ & Boltzmann equilibrium density \\
$H(x)$ & Hamiltonian of the phase point $x$ \\
$\beta=1/k_BT$ & Inverse temperature \\
$Z$ & Configurational partition function, $Z=\int dx\, e^{-\beta H(x)}$ \\
$\PP[\XP]$ & Free (unconditioned) path probability distribution \\
$A,\,B$ & Reactant / product stable states \\
$\mathbbm{1}_R(x)$, $\mathbbm{1}_A(x)$, $\mathbbm{1}_B(x)$ & Indicator function of region $R$ (or state $A$, $B$) \\
$\PP_{AB}[\XP]$ & Path distribution restricted to transition paths $A\to B$ \\
$\mathcal{Z}_{AB}$ & Normalization of $\PP_{AB}$ \\
\bottomrule
% \end{tabular}
% \end{table*}
% \begin{table*}[t]
% \centering
%\caption{
& TIS interfaces and crossing probabilities.\\
%\label{tab:defs-tis}
%\begin{tabular}{ll}
%\toprule
%Symbol & Description \\
\midrule
$\lambda$, $\lambda(x)$ & Order parameter / collective variable defining the interfaces \\
$\lambda_i$, $i=0,\dots,K$ & Location of interface $i$; 
$\lambda_0\equiv\lambda_A$, $\lambda_K\equiv\lambda_B$ \\
$K$ & Total number of interfaces \\
$\lambda_{\max}[\XP]$, $\Lambda$ & Maximum value of $\lambda$ reached along path $\XP$ \\
$\lambda_{\min}[\XP]$ & Minimum value of $\lambda$ reached along path $\XP$ \\
$\phi_A$ & Flux of trajectories leaving $A$ across the innermost interface $\lambda_0$ \\
$k_{AB}$, $k_{BA}$ & Forward / backward rate constant \\
$P_A(\lambda\,|\,\lambda_0)$ & Unconditional crossing probability of reaching $\lambda$, starting from $A$ \\
$P_A(\lambda_j\,|\,\lambda_i)$ & Conditional crossing probability from interface $i$ to interface $j$ \\
$\PP_A[\XP]$ & Path probability of excursions leaving $A$ (unbiased, before conditioning) \\
$\PP_{A,k}[\XP]$ & Path probability of the biased ensemble conditioned on crossing $\lambda_k$ \\
$N_k$ & Number of paths sampled in interface ensemble $k$ \\
$n$ & Running index over paths within an ensemble \\
\bottomrule
%\end{tabular}
%\end{table*}
%\begin{table*}[t]
%\centering
%\caption{
& Reweighted path ensemble (RPE) and weights.\\
%\label{tab:defs-weights}
%\begin{tabular}{ll}
%\toprule
%Symbol & Description \\
\midrule
$w_k$ & Weight assigned to interface (bin) $k$ \\
$w[\XP]$, $w_{k,n}$ & Weight assigned to an individual sampled path \\
$\mathcal{Z}=\sum_k\sum_n w_{k,n}$ & Normalization of the reweighted path ensemble \\
$g_k(\lambda)=\theta(\lambda_{k+1}-\lambda)\theta(\lambda-\lambda_k)$ & Bin-selector function (picks out interface bin $k$) \\
$h_k(\lambda)=\theta(\lambda-\lambda_k)$ & ``Has crossed interface $k$'' indicator (cumulative, one-sided) \\
$\mathcal{Z}_k$ & Partition function of ensemble $k$; $\mathcal{Z}_k\equiv P_A(\lambda_k|\lambda_0)$ \\
% $ Z_i=Z_i/\mathcal{Z}$ & Normalized partition function \\
%$\bar w_i$ & Single-sweep interface weight, $\bar w_i^{-1}=\sum_{k=0}^{i}N_k/ Z_k$ \\
$O[\XP]$, $\langle O\rangle$ & Path observable and its RPE-weighted average \\
\bottomrule
\end{tabular}
\end{table*}

\begin{table*}[h!]
\centering
\caption{Explanation of used symbols (continued)}
\label{tab:defs2}
\begin{tabular}{ll}
\toprule
Symbol & Description \\
\midrule
&{Path numbers, histograms, and the SAR.}\\
\midrule
$m_{A,k}(\lambda), m_{k}(\lambda)$ & Unnormalized crossing-probability histogram of ensemble $k$ \\
$m_A(\lambda)=\sum_k m_{A,k}(\lambda)$ & Summed (total) unnormalized histogram \\
$m_k$ & (Point matching only) number of paths of ensemble $k$ ending between $\lambda_k,\lambda_{k+1}$ \\
$n_i(\lambda_j)$ & Number of paths of ensemble $i$ reaching bin $j$ (``path number'') \\
$\tilde n(\lambda_k)$ & Reweighted path number, summed over ensembles \\
$D_j$ & Set of paths whose $\lambda_{\max}$ falls in bin $[\lambda_j,\lambda_{j+1})$ \\
$V_i=\sum_k n_k(\lambda_i)$ & Number of paths with $\lambda_{\max}\in[\lambda_i,\lambda_{i+1})$ (``deaths'' in bin $i$) \\
$S_i=\sum_k^i N_k$ & Cumulative number of paths started up to interface $i$ \\
$F_i=\sum_k^i V_k$ & Cumulative number of paths finished (died) up to interface $i$ \\
$s_A(\lambda_i)=S_i-F_i+V_i$ & Number of still-active (``surviving'') paths at $\lambda_i$ \\
$\lambda^{(i)}$, $i=0,\dots,M$ & Location of histogram bins $i$\\ 
$M$ & Number of histogram bins (Sec.~III.B; $M=K, \lambda^{(i)}=\lambda_i$ is the recommended choice) \\
\bottomrule
%\end{tabular}
%\end{table*}
%\begin{table*}[t]
%\centering
%\caption{Survival-analysis (hazard / product-limit) notation.}
%\label{tab:defs-survival}
%\begin{tabular}{ll}
%\toprule
%Symbol & Description \\
%\midrule
&Survival-analysis (hazard / product-limit) notation.\\
\midrule
$\Lambda_{(1)}<\Lambda_{(2)}<\dots<\Lambda_{(M)}$ & Ordered observed path maxima ($M$ = number of distinct maxima) \\
$q_i$ & Hazard: probability a path dies exactly at $\Lambda_{(i)}$, given survival to $\Lambda_{(i)}$ \\
$p_i$ & Survival probability at interface $i$ in the one-path-per-interface limit, $p_i=1-q_i$ \\
$d_i$ & Number of deaths at $\Lambda_{(i)}$ \\
$r_i$ & Size of the risk set at $\Lambda_{(i)}$ \\
$S(\lambda)$, $\hat S(\lambda)$ & Survival function (= crossing probability) and its product-limit estimator \\
$e_n$ & Entry point (interface) of path $n$ \\
$\Lambda_n$ & Death point (extreme value) of path $n$ \\
$i(n)$, $k(n)$ & Index of the death bin / entry ensemble for path $n$ \\
$W_n=\{j: e_n\le\Lambda_{(j)}<\Lambda_n\}$ & Observation window of path $n$ \\
\bottomrule
%\end{tabular}
%\end{table*}
%
%\begin{table*}[t]
%\centering
%\caption{Statistical synthesis: densities, mixtures, and likelihoods.}
%\label{tab:defs-synthesis}
%\begin{tabular}{ll}
%\toprule
%Symbol & Description \\
&Statistical synthesis: densities, mixtures, and likelihoods.\\
\midrule
$p(\Lambda)$ & Probability density of path maxima (``density of path numbers'') \\
$p_k(\Lambda)$ & Marginal density of maxima within biased ensemble $k$ \\
$\PA[\XP\,|\,\Lambda]$ & Path distribution conditioned on $\lambda_{\max}[\XP]=\Lambda$ \\
$q[\XP]$ & Merged (mixture) path distribution over all ensembles \\
%$\vartheta=(p,\mathcal{C})$ & Nonparametric model parameters \\
%$\mathcal{C}=\{\PA[\cdot\,|\,\Lambda]\}$ & Collection of conditional path measures \\
$L$, $\mathcal{L}_A$ & Likelihood functions \\
$\alpha_i$ & Combination weight in the variance-based estimator \\
$b_i$ & Scaling factor relating a biased and the unbiased ensemble average \\
$\epsilon_i$ & Relative statistical error of the ensemble-$i$ estimate (VBR method) \\
$c$ & Statistical inefficiency / correlation time between sampled paths \\
\end{tabular}
\end{table*}

\bibliographystyle{apsrev4-2}
\bibliography{references}

\end{document}